\documentclass{elsart}
\journal{Physics Letter B}
\usepackage{graphics}

\makeatletter
\DeclareRobustCommand*\cal{\@fontswitch\relax\mathcal}
\DeclareRobustCommand*\mit{\@fontswitch\relax\mathnormal}

\def\endequation{\eqno \hbox{\@eqnnum}$$\@ignoretrue}
\def\eqnarray{%
   \stepcounter{equation}%
   \def\@currentlabel{\p@equation\theequation}%
   \global\@eqnswtrue
   \m@th
   \global\@eqcnt\z@
   \tabskip\@centering
   \let\\\@eqncr
   $$\everycr{}\halign to\displaywidth\bgroup
       \hskip\@centering$\displaystyle\tabskip\z@skip{##}$\@eqnsel
      &\global\@eqcnt\@ne\hskip \tw@\arraycolsep \hfil${##}$\hfil
      &\global\@eqcnt\tw@ \hskip \tw@\arraycolsep
         $\displaystyle{##}$\hfil\tabskip\@centering
      &\global\@eqcnt\thr@@ \hb@xt@\z@\bgroup\hss##\egroup
         \tabskip\z@skip
      \cr
}
\def\endeqnarray{%
      \@@eqncr
      \egroup
      \global\advance\c@equation\m@ne
   $$\@ignoretrue
}
\makeatother

\def\Vec#1{\mbox{\boldmath $#1$}}
\def\Cdot{\!\cdot\!}

\begin{document}

\begin{frontmatter}

\title{
Search for the Electric Dipole Moment of the $\tau$ Lepton
}

\collab{Belle Collaboration}
  \author[Nagoya]{K.~Inami}, 
  \author[KEK]{K.~Abe}, 
  \author[TohokuGakuin]{K.~Abe}, 
  \author[Niigata]{R.~Abe}, 
  \author[Tohoku]{T.~Abe}, 
  \author[KEK]{I.~Adachi}, 
  \author[Tokyo]{H.~Aihara}, 
  \author[Nagoya]{M.~Akatsu}, 
  \author[Tsukuba]{Y.~Asano}, 
  \author[Toyama]{T.~Aso}, 
% ...

% ...
\address[KEK]{High Energy Accelerator Research Organization (KEK), Tsukuba, Japan}
% ...
\address[Nagoya]{Nagoya University, Nagoya, Japan}
% ...
\address[Niigata]{Niigata University, Niigata, Japan}
% ...
\address[TohokuGakuin]{Tohoku Gakuin University, Tagajo, Japan}
\address[Tohoku]{Tohoku University, Sendai, Japan}
\address[Tokyo]{University of Tokyo, Tokyo, Japan}
% ...
\address[Toyama]{Toyama National College of Maritime Technology, Toyama, Japan}
\address[Tsukuba]{University of Tsukuba, Tsukuba, Japan}
% ...
\thanks[NovaGorica]{on leave from Nova Gorica Polytechnic, Nova Gorica, Slovenia}

\begin{abstract}
We have searched for a CP violation signature arising from
an electric dipole moment~($d_\tau$) of the $\tau$ lepton
in the $e^+e^- \to \tau^+\tau^-$ reaction.
Using an optimal observable method
and 29.5 fb$^{-1}$ of data collected with the Belle detector 
at the KEKB collider at $\sqrt{s}=10.58$~GeV,
we find
$Re(d_\tau) = ( 1.15 \pm 1.70 ) \times 10^{-17} e\,{\rm cm}$ and
$Im(d_\tau) = ( -0.83 \pm 0.86 ) \times 10^{-17} e\,{\rm cm}$
and set the 95\% confidence level limits
$-2.2 < Re(d_\tau) < 4.5 ~(10^{-17}e\,{\rm cm})$ and
$-2.5 < Im(d_\tau) < 0.8 ~(10^{-17}e\,{\rm cm})$.

\end{abstract}
\begin{keyword}
tau \sep electic dipole moment \sep CP violation \sep optimal observable
\PACS 13.40.Gp \sep 13.35.Dx \sep 14.60.Fg
\end{keyword}

\end{frontmatter}

\clearpage

\section{Introduction}

While large CP violating asymmetry has recently been confirmed in B-meson
decay~\cite{ref:BCPV1,ref:BCPV2}, 
the Standard Model (SM) does not predict any appreciable CP
violation (CPV) in the lepton sector. However, physics beyond the SM could
produce CPV in leptonic processes; we would expect such effects to be
enhanced for $\tau$ leptons due to their large mass. Parameterizing CPV in
$\tau$-pair production by an electric dipole moment $d_\tau$, several authors
have found possible effects of order $|d_\tau| \sim 10^{-19} e$\,cm due to new
physics models~\cite{ref:th1,ref:th2}. 
The best existing bounds on $d_\tau$ are indirect,
requiring $|d_\tau| < {\rm O}(10^{-17}) e$\,cm 
based on naturalness arguments~\cite{ref:th3} and precision LEP 
data~\cite{ref:th4}.\footnote{A very strict constraint 
$|d_\tau| < 2.2 \times 10^{-25} e$\,cm
has also been derived from the experimental limits on 
$\mu \to e \gamma$ decay~\cite{ref:th5}. We note, however,
that this result assumes a particular ansatz for the lepton mixing
matrix. In a recent preprint~\cite{ref:th6}, other authors have argued that
the same constraint may be derived under weaker assumptions.}
Previous direct
measurements have been performed at LEP, where L3~\cite{ref:L3} found 
$-3.1 < Re(d_\tau) < 3.1 \times 10^{-16} e$\,cm and 
OPAL~\cite{ref:OPAL} found $|Re(d_\tau)| < 3.7 \times 10^{-16} e$\,cm, 
using the process $e^+e^- \to \tau^+ \tau^- \gamma$; and 
by ARGUS~\cite{ref:ARGUS}, which set limits 
$|Re(d_\tau)| < 4.6 \times 10^{-16} e$\,cm and 
$|Im(d_\tau)| < 1.8 \times 10^{-16} e$\,cm 
based on a study of $e^+e^- \to \tau^+ \tau^-$ production. 

In this paper we present the first direct measurement of the $\tau$ lepton's
electric dipole moment with a sensitivity in the $10^{-17} e$\,cm range.
We improve on the
sensitivity of ARGUS by an order of magnitude, by reducing the systematic
uncertainty in the extraction of $d_\tau$, and by analyzing a much larger
sample of events.

We search for CP violating effects at the $\gamma\tau\tau$ vertex 
in the process $e^+e^- \to \gamma^* \to \tau^+ \tau^-$
using triple momentum and spin correlation observables. 
The CP violating effective Lagrangian can be expressed as
\begin{equation}
{\cal L}_{CP} = - i d_\tau(s) \bar{\tau} \sigma^{\mu \nu} \gamma_5 \tau
  \partial_\mu A_\nu, \label{eq:Lagrangian}
\end{equation}
where the electric dipole form factor $d_\tau$ depends in general on $s$, 
the squared energy of the $\tau$-pair system. 
In common with other authors, we ignore this possible $s$-dependence,
assuming $d_\tau(s) \equiv d_\tau$, which is constant.
($d_\tau(0)$ corresponds to the electric dipole moment of the $\tau$, 
and we shall use this term hereafter.)
The squared spin density matrix (${\cal M}^2_{\rm prod}$) 
for $\tau$-pair production
$e^+(\Vec{p})e^-(-\Vec{p}) \to \tau^+(\Vec{k},\Vec{S}_+) \tau^-(-\Vec{k},
\Vec{S}_-)$ is given by~\cite{ref:EDM}
\small
\begin{eqnarray}
{\cal M}_{\rm prod}^2 &=& {\cal M}_{\rm SM}^2 
 + Re(d_\tau) {\cal M}_{Re}^2
 + Im(d_\tau) {\cal M}_{Im}^2
 + |d_\tau|^2 {\cal M}_{d^2}^2 , \\
{\cal M}_{\rm SM}^2 &=& \frac{e^4}{k_0^2}
[   k_0^2 + m_\tau^2 + |\Vec{k}^2|(\hat{\Vec{k}}\Cdot\hat{\Vec{p}})^2 
 - \Vec{S}_+ \Cdot \Vec{S}_- |\Vec{k}|^2 
(1-(\hat{\Vec{k}}\Cdot\hat{\Vec{p}})^2)
\nonumber \\ & &
 + 2(\hat{\Vec{k}}\Cdot\Vec{S}_+)(\hat{\Vec{k}}\Cdot\Vec{S}_-)
    (|\Vec{k}|^2+(k_0 - m_\tau)^2 (\hat{\Vec{k}}\Cdot\hat{\Vec{p}})^2)
 + 2 k_0^2 (\hat{\Vec{p}}\Cdot\Vec{S}_+) (\hat{\Vec{p}}\Cdot\Vec{S}_-)
\nonumber \\ & &
 - 2 k_0(k_0-m_\tau)(\hat{\Vec{k}}\Cdot\hat{\Vec{p}})
    ((\hat{\Vec{k}}\Cdot\Vec{S}_+)(\hat{\Vec{p}}\Cdot\Vec{S}_-) + 
     (\hat{\Vec{k}}\Cdot\Vec{S}_-)(\hat{\Vec{p}}\Cdot\Vec{S}_+) ) ],
\\
{\cal M}_{Re}^2 &=& 4 \frac{e^3}{k_0} |\Vec{k}| [
 - (m_\tau + (k_0-m_\tau)(\hat{\Vec{k}}\Cdot\hat{\Vec{p}})^2)
   (\Vec{S}_+ \!\times\! \Vec{S}_-)\Cdot\hat{\Vec{k}}
 + k_0 (\hat{\Vec{k}}\Cdot\hat{\Vec{p}})
   (\Vec{S}_+ \!\times\! \Vec{S}_-)\Cdot\hat{\Vec{p}}
] , ~~~~
\\
{\cal M}_{Im}^2 &=& 4 \frac{e^3}{k_0} |\Vec{k}| [
 - (m_\tau + (k_0-m_\tau)(\hat{\Vec{k}}\Cdot\hat{\Vec{p}})^2)
   (\Vec{S}_+ \!-\! \Vec{S}_-)\Cdot\hat{\Vec{k}}
 + k_0 (\hat{\Vec{k}}\Cdot\hat{\Vec{p}})
   (\Vec{S}_+ \!-\! \Vec{S}_-)\Cdot\hat{\Vec{p}}
] , ~~~~
\\
{\cal M}_{d^2}^2 &=& 4 e^2 |\Vec{k}|^2 \Cdot 
 (1-(\hat{\Vec{k}}\Cdot\hat{\Vec{p}})^2) (1-\Vec{S}_+ \Cdot \Vec{S}_-),
\end{eqnarray}
\normalsize
where $k_0$ is the energy of the $\tau$, $m_\tau$ is the $\tau$ mass, 
$\Vec{p}$ is the momentum vector of $e^+$, $\Vec{k}$ is 
the momentum vector of $\tau^+$ in the center-of-mass frame,
$\Vec{S}_\pm$ are the spin vectors for $\tau^\pm$, and the hat denotes 
a unit momentum.
We disregard the higher order terms proportional to $|d_\tau^2|$,
which is valid since $d_\tau$ is small.
${\cal M}_{\rm SM}^2$ corresponds to the SM term.
${\cal M}_{Re}^2$ and ${\cal M}_{Im}^2$ are the interference terms 
(related to the real and imaginary parts of $d_\tau$, respectively)
between the SM and CPV amplitudes.
${\cal M}_{Re}^2$ is CP odd and T odd, while ${\cal M}_{Im}^2$ is CP
odd, but T even.
In the above equations, $e^+$ and $e^-$ are assumed to be unpolarized and
massless particles.

We adapt the so-called optimal observable method~\cite{ref:Optimal}, which 
maximizes the sensitivity to $d_\tau$.
Here the optimal observables are defined as
\begin{equation}
{\cal O}_{Re} = \frac{{\cal M}_{Re}^2}{{\cal M}_{\rm SM}^2},~~~
{\cal O}_{Im} = \frac{{\cal M}_{Im}^2}{{\cal M}_{\rm SM}^2}.
\end{equation}
The mean value of the observable ${\cal O}_{Re}$ is given by
\begin{equation}
\langle{\cal {O}}_{Re}\rangle \propto
\int {\cal O}_{Re} {\cal M}^2_{\rm prod} d\phi
= \int {\cal{M}}^2_{Re} d\phi + 
Re(d_{\tau}) \int \frac{({\cal {M}}^2_{Re})^2}
{{\cal {M}}^2_{\rm{SM}}} d\phi,
\label{eq:obs}
\end{equation}
where the integration is over the phase space ($\phi$) spanned by 
the relevant kinematic variables.
The cross-term containing the integral of the product of
${\cal M}^2_{Re}$ and ${\cal M}^2_{Im}$ drops out 
because of their different symmetry properties.
The expression for the imaginary part is similar.
The means of the observables
$\langle {\cal O}_{Re} \rangle$ and $\langle {\cal O}_{Im} \rangle$ 
are therefore linear functions of $d_{\tau}$,
\begin{equation}
 \langle{\cal O}_{Re}\rangle =
   a_{Re} \cdot Re(d_\tau) +b_{Re},~~~
 \langle{\cal O}_{Im}\rangle =
   a_{Im} \cdot Im(d_\tau) +b_{Im}.
 \label{eq:relation1}
\end{equation}

Eight different final states in the decay of $\tau$-pairs,
$(e\nu\bar{\nu})(\mu\nu\bar{\nu})$, $(e\nu\bar{\nu})(\pi\nu)$, 
$(\mu\nu\bar{\nu})(\pi\nu)$, $(e\nu\bar{\nu})(\rho\nu)$, 
$(\mu\nu\bar{\nu})(\rho\nu)$, $(\pi\nu)(\rho\nu)$, 
$(\rho\nu)(\rho\bar{\nu})$, and $(\pi\nu)(\pi\bar{\nu})$, are analyzed,
where all particles except $\nu$ and $\bar{\nu}$ are positively or negatively 
charged.
Because of the undetectable particles, we can not fully reconstruct
the quantities $\Vec{k}$ and $\Vec{S}_\pm$. 
Therefore, for each event we calculate possible kinematic 
configurations and obtain 
the mean value of ${\cal M}^2_{\rm SM}$, 
${\cal M}^2_{Re}$ and ${\cal M}^2_{Im}$
by averaging over the calculated configurations.
In the case when both $\tau$ leptons decay hadronically 
($\tau \to \pi\nu$ or $\rho\nu$),
the $\tau$ flight direction is calculated with a two-fold 
ambiguity~\cite{ref:taucalc}
and we take the average of ${\cal M}^2_{\rm SM}$, 
${\cal M}^2_{Re}$ and ${\cal M}^2_{Im}$ over the two solutions.
In the case when either one or both $\tau$ leptons decay leptonically
($\tau \to e\nu\bar{\nu}$ or $\mu\nu\bar{\nu}$), 
a Monte Carlo (MC) treatment is adopted to take into account
the additional ambiguity in the effective mass of the $\nu\bar{\nu}$ system 
($m_{\nu\bar{\nu}}$). 
For each event we generate 100 MC configurations using a hit-and-miss approach
by varying $m_{\nu\bar{\nu}}$, and compute the averaged 
${\cal M}^2_{\rm SM}$, ${\cal M}^2_{Re}$ and ${\cal M}^2_{Im}$
over successful tries
in which the $\tau$ direction can be constructed kinematically. 
In the calculation,
we ignore the effect of undetected photons coming from initial
state radiation, radiative $\tau$ decays and bremsstrahlung.
The resulting bias in $d_\tau$ is included among the systematic errors;
the effect on the final result is negligible.

\section{Data and event selection}

In this analysis, we use 26.8 million $\tau$-pairs ($29.5~{\rm fb}^{-1}$)
accumulated with 
the Belle detector~\cite{ref:Belle} at the KEKB accelerator~\cite{ref:KEKB}.
KEKB is an asymmetric energy $e^+e^-$ collider
with a beam crossing angle of 22 mrad. Its center-of-mass energy is
10.58 GeV, corresponding to the $\Upsilon(4S)$ resonance, 
with beam energies of 8 and 3.5 GeV for electrons and positrons,
respectively.
Belle is a general-purpose detector with an asymmetric structure 
along the beam direction. Among the detector elements, 
the central drift chamber (CDC) and the silicon vertex detector (SVD)
are essential to obtain the momentum vectors of charged particles.
The combined information from the silica Aerogel Cherenkov counters (ACC),
the time-of-flight counters (TOF), the CsI electromagnetic calorimeter (ECL), 
and the $\mu/K_L$ detector (KLM) is used for particle identification.

The MC event generators
KORALB/TAUOLA~\cite{ref:KORALB} are used for 
$\tau$-pair production and decays.
The detector simulation is performed with a GEANT-based program, GSIM.
Actual data and MC generated events are reconstructed 
by the same program written by the Belle collaboration. 

\begin{table}
\renewcommand {\baselinestretch}{0.75}
 \caption{Yield, purity and background rate obtained for the event selection
described in the text, where the purity is evaluated by MC
simulation and its error comes from MC statistics. }
 \label{table:selection.result}
 \begin{center}
 \begin{tabular}{crcl}
  \hline
       & Yield    & Purity (\%)     & Background mode (\%)\\
  \hline
  $e\mu$     & 250,948    & $96.6\pm0.1$ & $2\gamma \to \mu\mu$(1.9), $\tau\tau \to e\pi$(1.1). \\
  $e\pi$     & 132,574    & $82.5\pm0.1$ & $\tau\tau \to e\rho$(6.0), $eK$(5.4), $e\mu$(3.1), $eK^*$(1.3). \\
  $\mu\pi$   & 123,520    & $80.6\pm0.1$ & $\tau\tau \to \mu\rho$(5.7), $\mu K$(5.3), $\mu\mu$(2.9), $2\gamma \to \mu\mu$(2.0). \\
  $e\rho$    & 240,501    & $92.4\pm0.1$ & $\tau\tau \to e\pi\pi^0\pi^0$(4.4), $eK^*$(1.7). \\
  $\mu\rho$  & 217,156    & $91.6\pm0.1$ & $\tau\tau \to \mu\pi\pi^0\pi^0$(4.2), $\mu K^*$(1.6), $\pi\rho$(1.0). \\
  $\pi\rho$  & 110,414    & $77.7\pm0.1$ & $\tau\tau \to \rho\rho$(5.1), $K \rho$(4.9), $\pi\pi\pi^0\pi^0$(3.8), $\mu\rho$(2.7). \\
  $\rho\rho$ & 93,016     & $86.2\pm0.1$ & $\tau\tau \to \rho\pi\pi^0\pi^0$(8.0), $\rho K^*$(3.1). \\
  $\pi\pi$   & 28,348     & $70.0\pm0.2$ & $\tau\tau \to \pi\rho$(9.2), $\pi K$(9.2), $\pi\mu$(4.7), $\pi K^*$(2.0). \\
  \hline
 \end{tabular}
 \end{center}
\end{table}

We reconstruct the eight $\tau$-pair decay modes mentioned above,
with the following conditions.
Each charged track is required to have a transverse momentum $p_t>0.1$ 
GeV/$c$. 
Photon candidates should deposit an energy of $E>0.1$ GeV 
in the ECL.
A signal event is required to have two charged tracks with zero net-charge 
and no photon apart from $\rho^\pm \to  \pi^\pm \pi^0$, 
$\pi^0 \to \gamma \gamma$.

A track is identified as an electron using a likelihood ratio
combining $dE/dx$ in the CDC,
the ratio of energy deposited in the ECL and momentum
measured in the CDC,
the shower shape in the ECL
and the hit pattern from the ACC.
The identification efficiency is estimated to be 92\% 
with a $\pi^\pm$ fake rate of 0.3\%
for the momentum range between 1.0 GeV/$c$ and 3.0 GeV/$c$
in the laboratory frame~\cite{ref:eID}.
A muon is identified by its range and hit pattern in the KLM detector,
with efficiency and fake rate estimated to be 91\% and 2\%,
respectively, for momenta in the laboratory frame greater than 1.2 GeV/$c$.
A track is considered to be a pion if it is identified as a hadron
by the KLM information and not identified as an electron:
the efficiency of this selection is estimated to be 81\%, and
the purity for the selected samples is estimated by MC to be 89\%.
A $\rho^\pm$ is reconstructed from a charged track and a $\pi^0$ where 
the track should be neither an electron nor a muon,
and for $\pi^0 \to \gamma\gamma$, the reconstructed $\pi^0$ should have an 
invariant mass between 110 and 150 MeV/$c^2$ and 
a momentum in the laboratory frame larger than $0.2$ GeV/$c$.
In order to suppress background
and to improve the performance of the particle identification cuts,
we restrict the analysis to lepton candidates within
the barrel region, $-0.60<\cos \theta< 0.83$,
and to pion candidates for $\tau \to \pi \nu$ within the KLM barrel region,
$-0.50<\cos \theta<0.62$,
where $\theta$ is the polar angle opposite to the $e^+$ beam direction
in the laboratory frame.
For the same reason, the laboratory frame particle momentum 
is required to be greater 
than 0.5 GeV/$c$ for an electron, 1.2 GeV/$c$ for both a muon and pion,
and 1.0 GeV/$c$ for the $\rho^\pm$.

The dominant backgrounds are due to two-photon as well as Bhabha and 
$\mu\mu$ processes. 
In order to remove two-photon events,
we require the missing momentum not to be directed towards the beam-pipe
region 
(imposing a selection $-0.950 <\cos \theta<0.985$),
and to reject the latter processes we require
that the sum of the charged track momenta
be less than 9 GeV/$c$ in the center-of-mass frame.
Additional selections are imposed particularly on the $e\pi$ mode 
where a large number of Bhabha events could contribute
through misidentification. For the $e\pi$ mode,
we remove events which satisfy the following criteria:
the opening angle of the two tracks in the plane perpendicular to the beam axis
is greater than $175^\circ$, and their momentum sum is greater
than 6 GeV/$c$ in the $\tau$-pair rest frame.
Finally, we remove events in which
the $\tau$ flight direction cannot be kinematically reconstructed,
which mostly arise from $\tau$-pairs 
having hard initial-state radiation 
and misidentified $\tau$-pair backgrounds.

The yield of events passing this selection is given in 
Table~\ref{table:selection.result} for each of the eight selected modes.
The mean energy of the $\tau$-pair system in the obtained sample
is $\sqrt{s}=10.38$~GeV; this sets the scale at which $d_\tau(s)$ is measured.
Because of events with soft radiated photons,
the energy scale is slightly lower than the beam energy.
The dominant background sources are also listed in the table.
Hadronic $\tau$ decays with two or more $\pi^0$'s make a contribution
of a few percent.
For the modes including $\pi^\pm$, kaons and misidentified muons
from other $\tau$ decays produce a feed-across background;
for example, we estimate that 5.3\% of the $\mu \pi$ sample consists of
$\mu K$ decays.
The other backgrounds are estimated by MC to be a few percent
from two-photon processes,
and less than 1\% from Bhabha, $\mu\mu$, and multihadronic processes.

\begin{figure}[t]
\centerline{\resizebox{4.5cm}{4.5cm}{\includegraphics{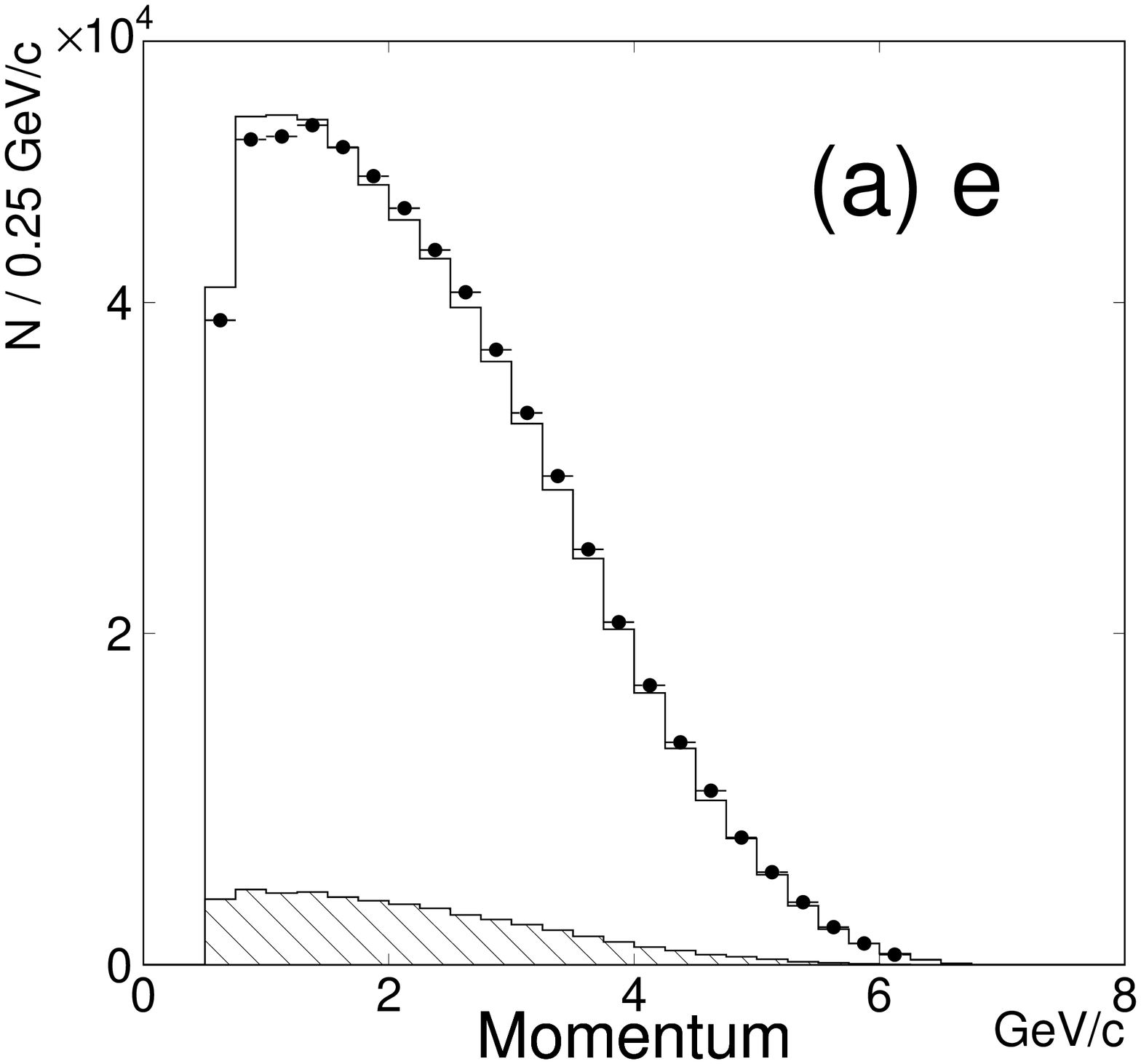}}
            \resizebox{4.5cm}{4.5cm}{\includegraphics{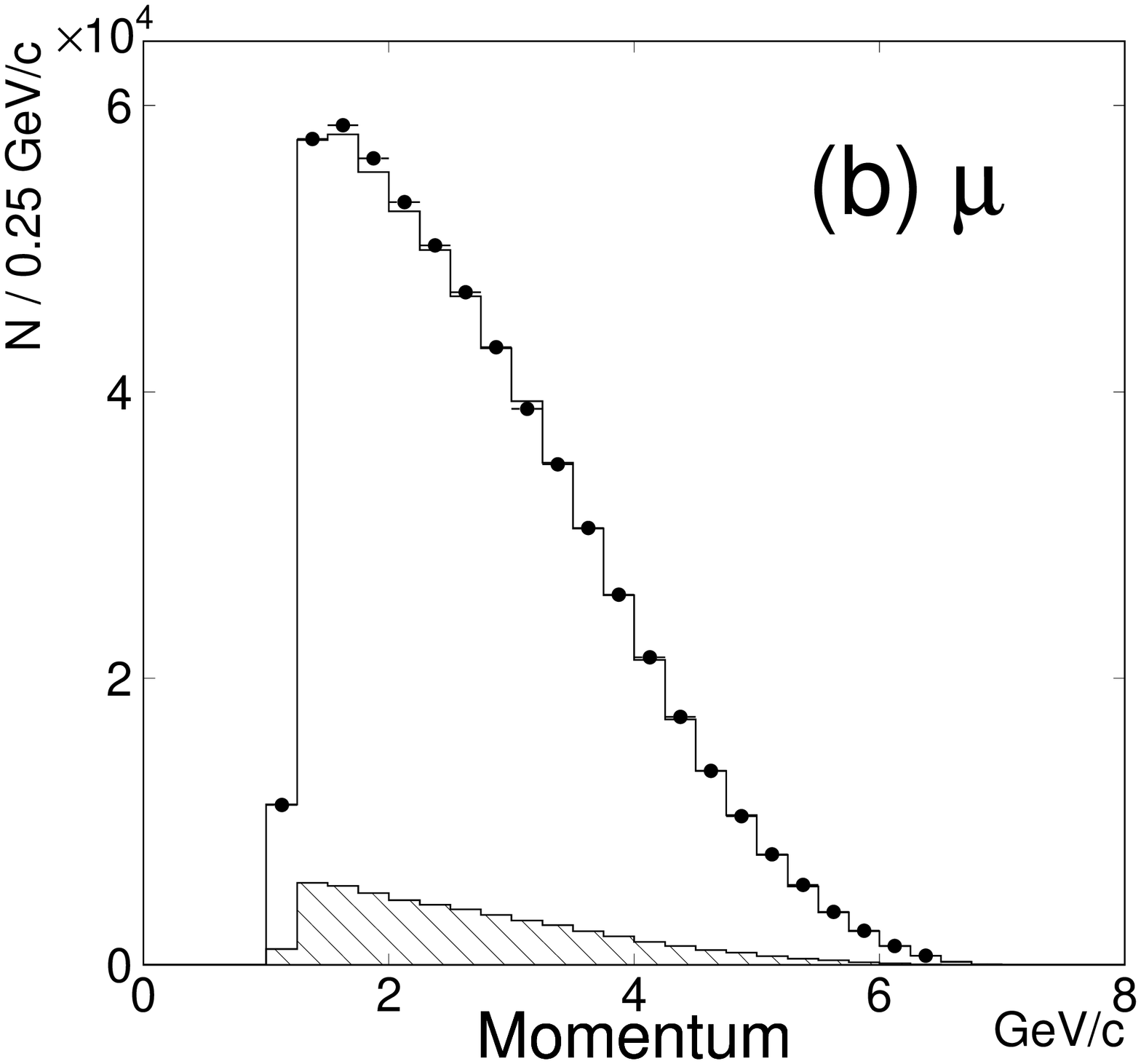}}}
\centerline{\resizebox{4.5cm}{4.5cm}{\includegraphics{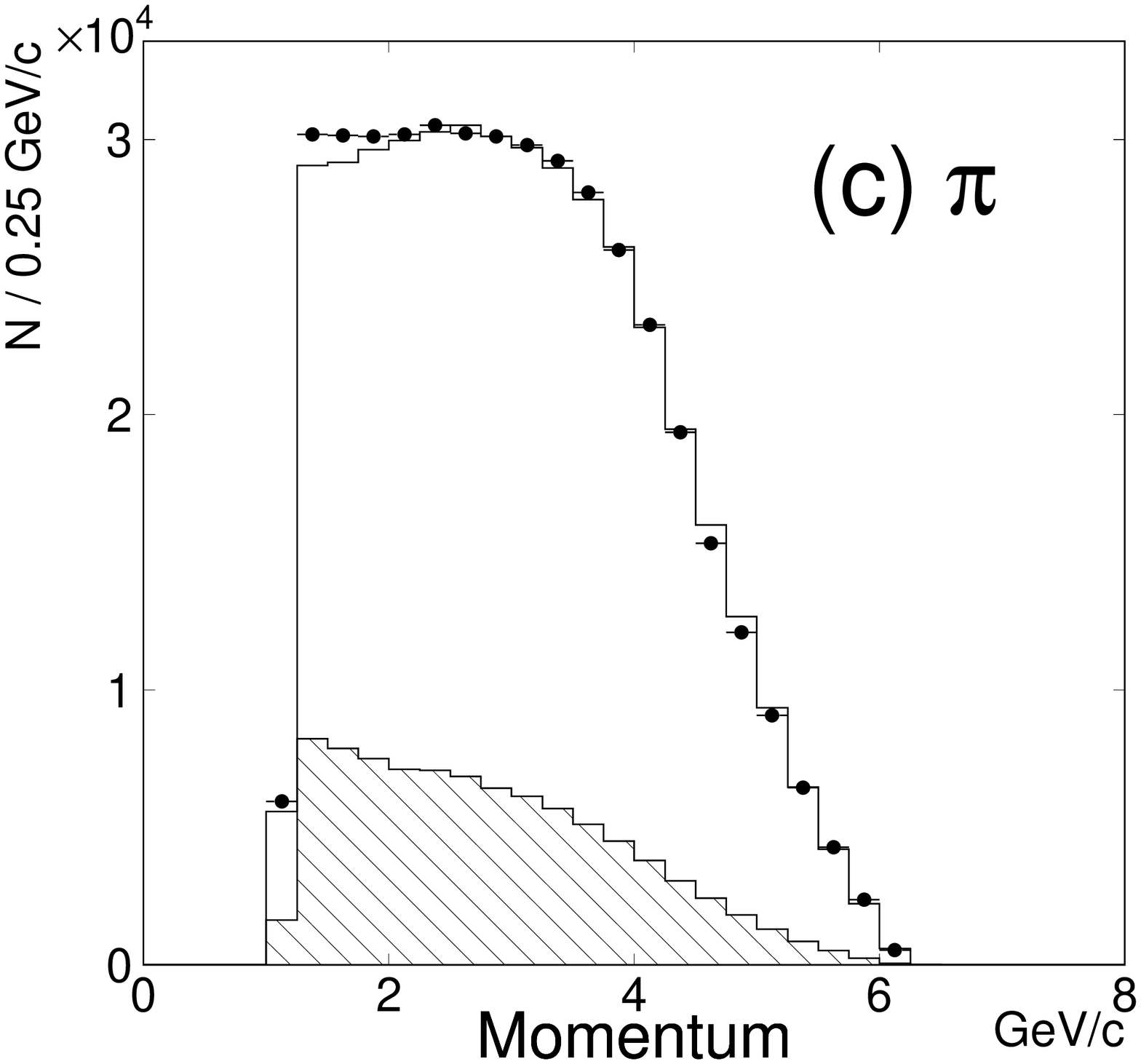}}
            \resizebox{4.5cm}{4.5cm}{\includegraphics{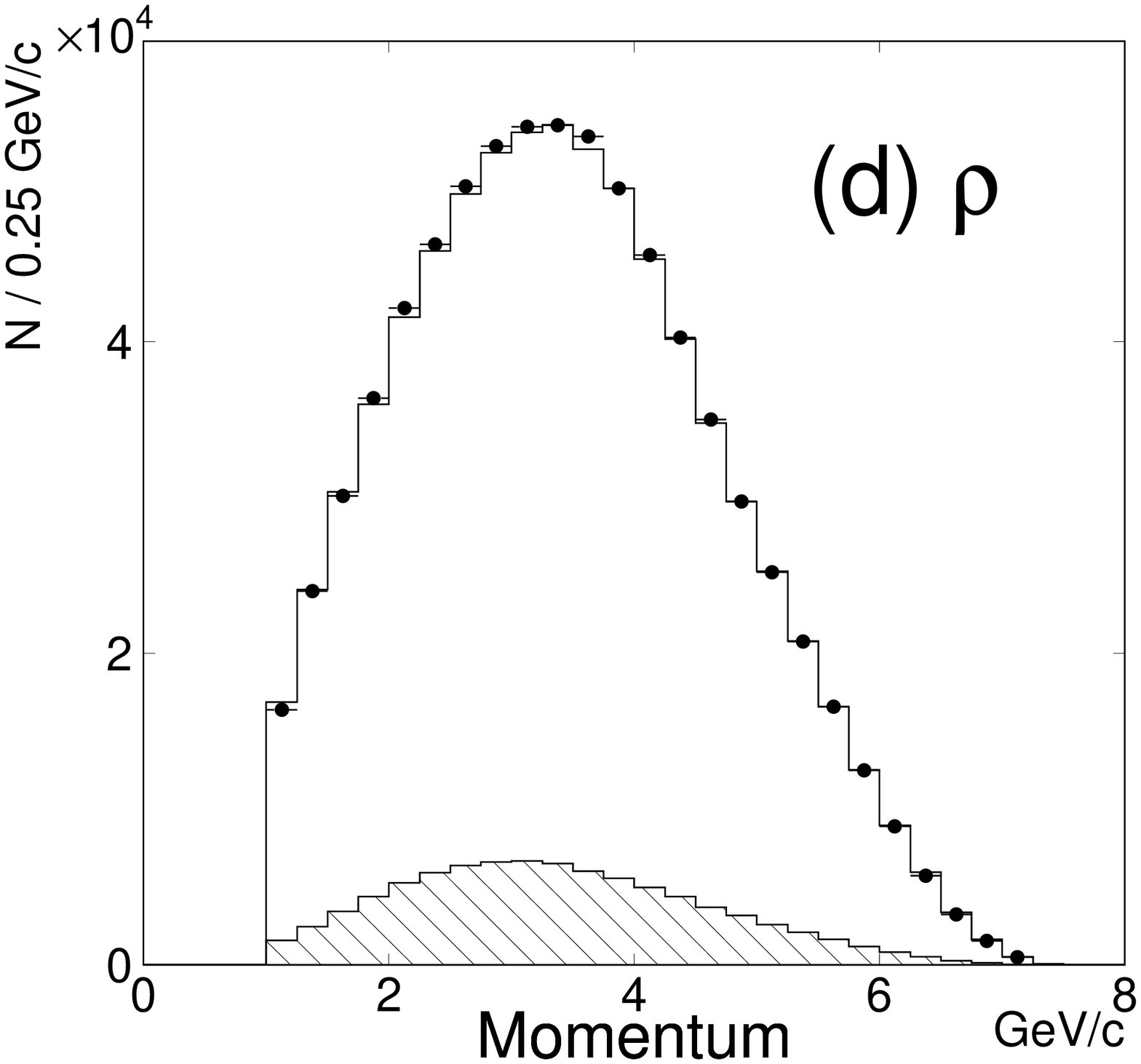}}}
\renewcommand {\baselinestretch}{0.75}
\caption{Momentum distributions of 
(a) $e^\pm$, (b) $\mu^\pm$, (c) $\pi^\pm$, and (d) $\rho^\pm$ in 
the laboratory frame.
The points with error bars are the data and the histogram is the MC 
expectation. 
The latter is scaled to the total number of entries.
The hatched histogram is the background distribution evaluated by MC.}
\label{fig:mom}
\end{figure}

\begin{figure}
\centerline{\resizebox{4.5cm}{4.5cm}{\includegraphics{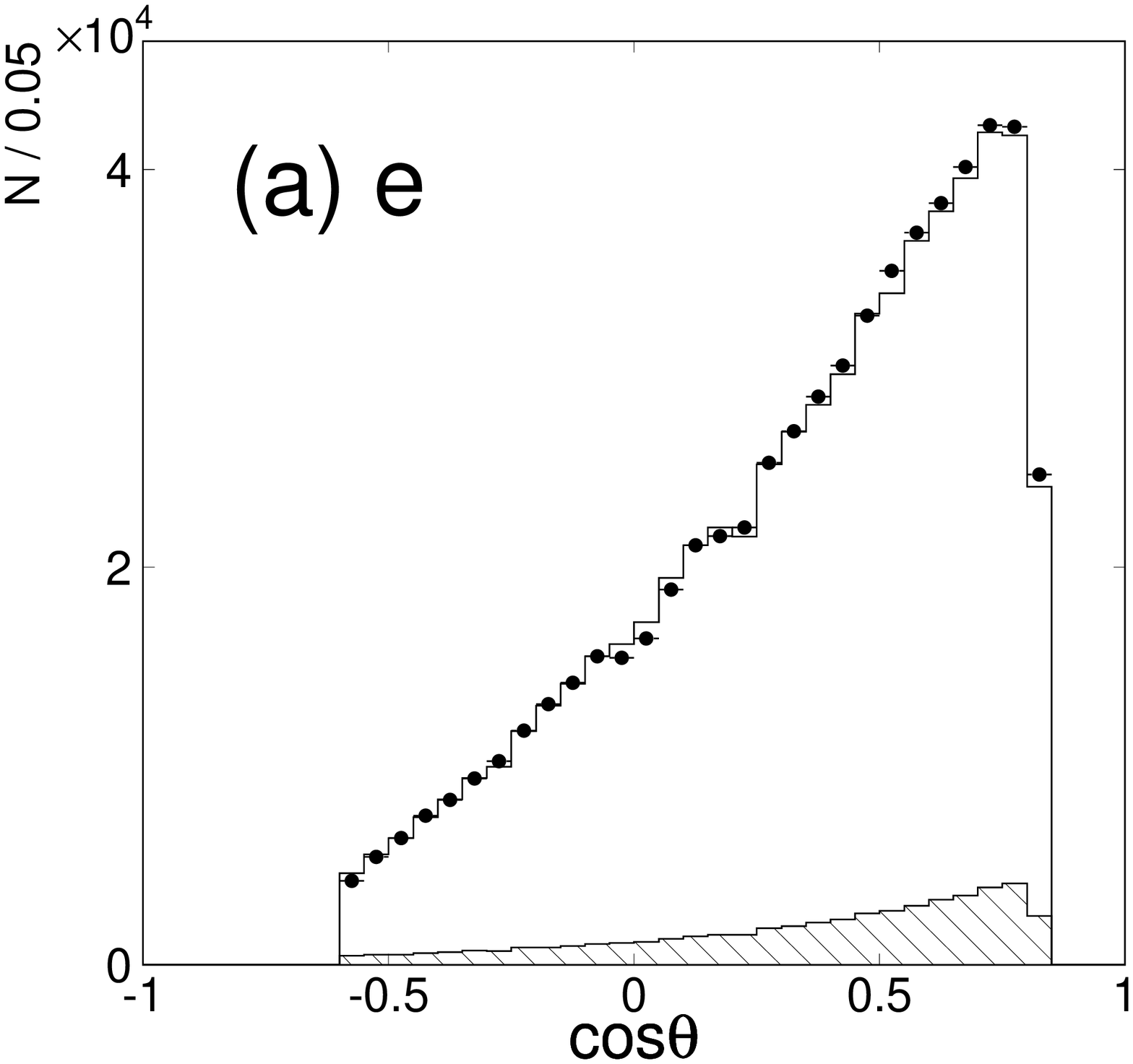}}
            \resizebox{4.5cm}{4.5cm}{\includegraphics{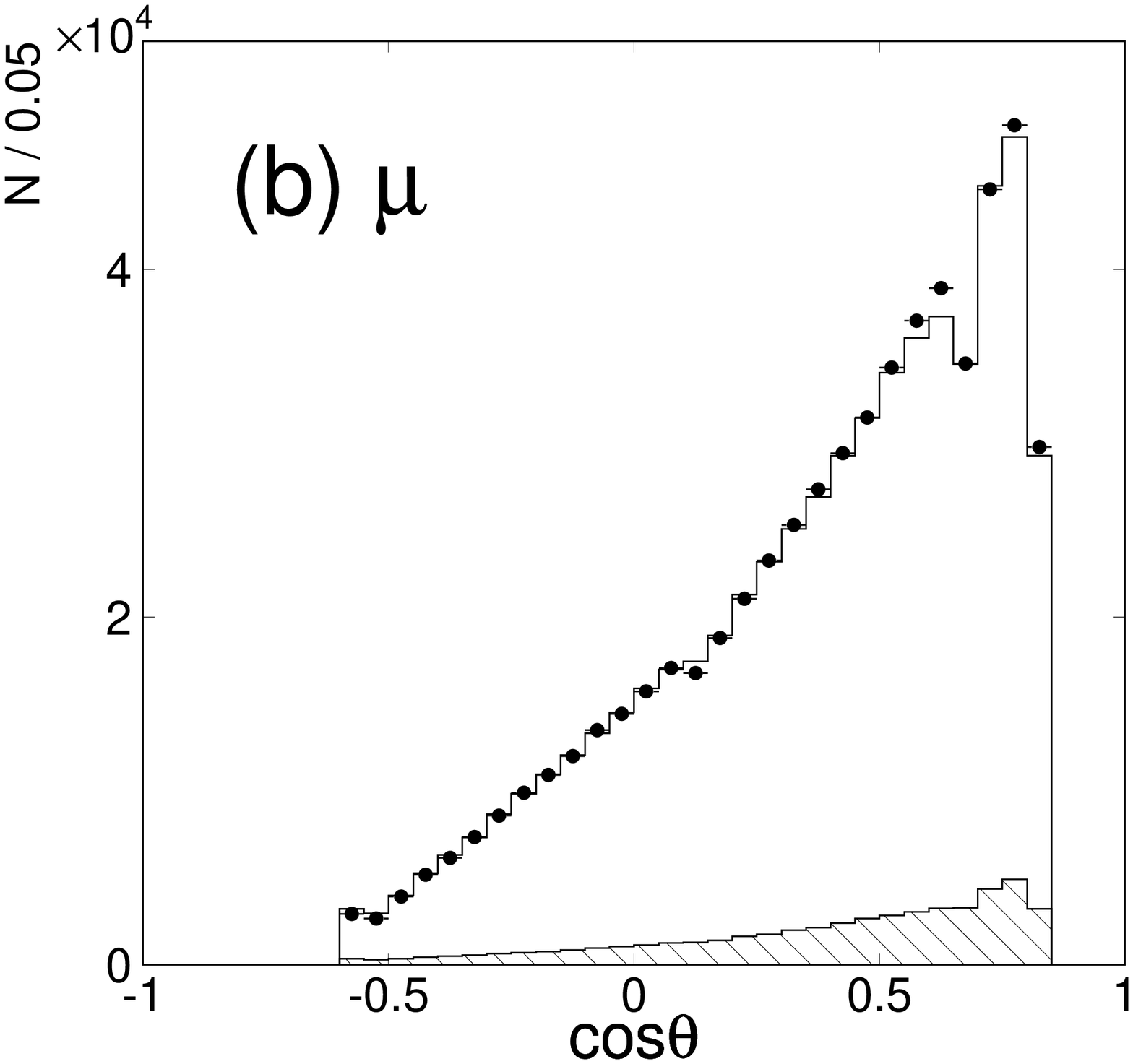}}}
\centerline{\resizebox{4.5cm}{4.5cm}{\includegraphics{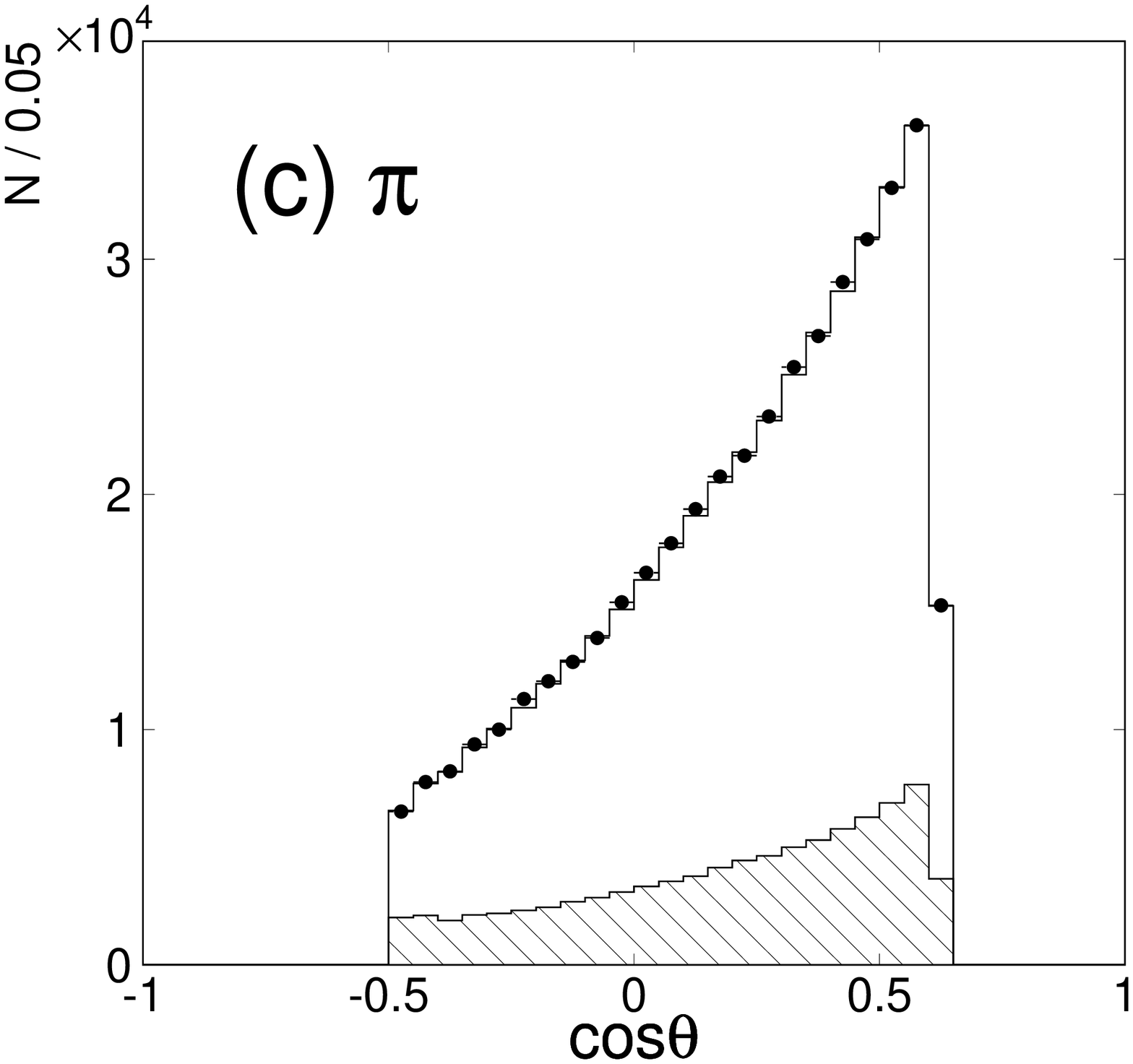}}
            \resizebox{4.5cm}{4.5cm}{\includegraphics{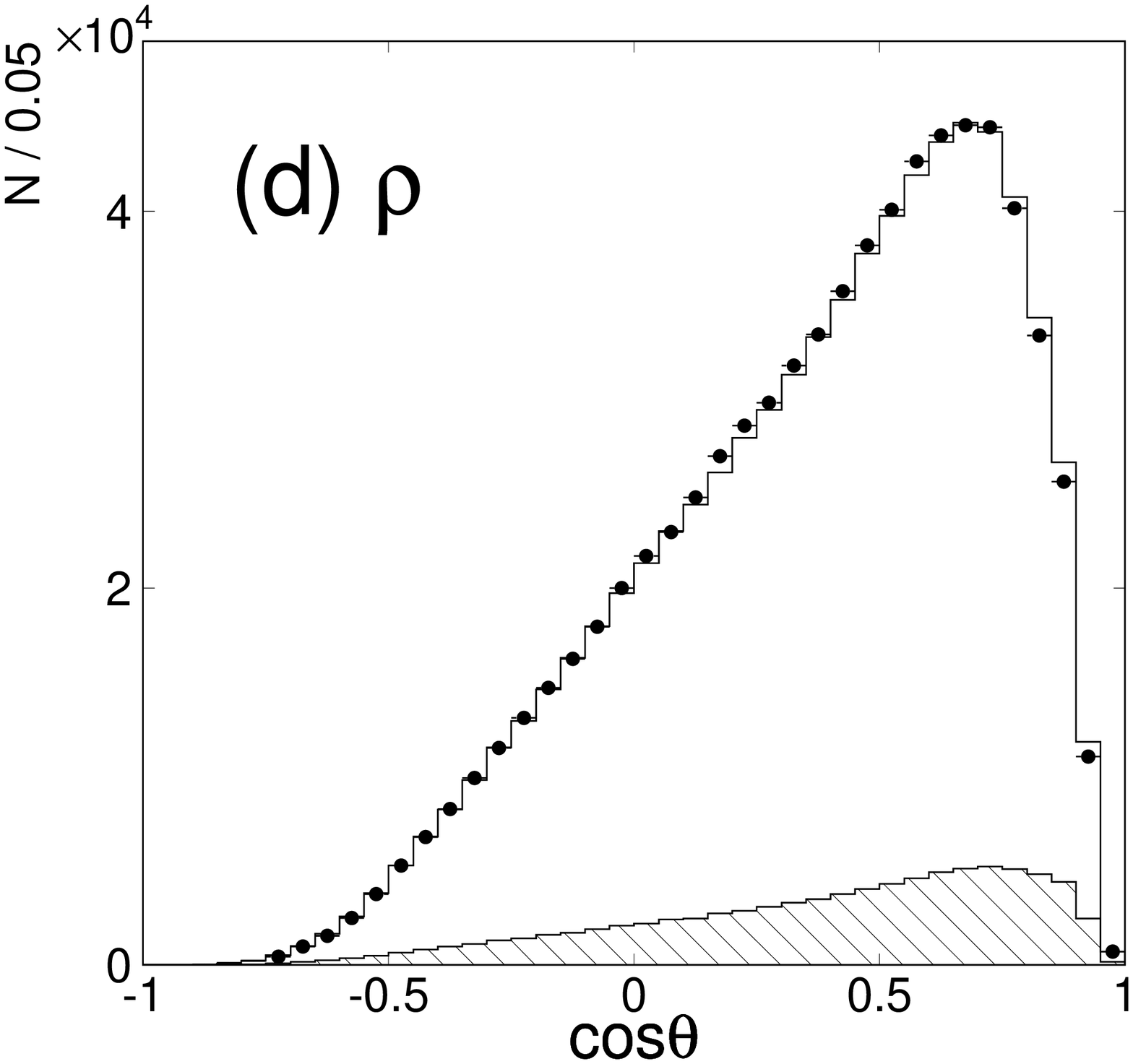}}}
\renewcommand {\baselinestretch}{0.75}
\caption{The $\cos \theta$ distributions of
(a) $e^\pm$, (b) $\mu^\pm$, (c) $\pi^\pm$, and (d) $\rho^\pm$ in 
the laboratory frame.
The meanings of the points and histograms are the same as 
in Fig.~\ref{fig:mom}.}
\label{fig:cos}
\end{figure}

Figures~\ref{fig:mom} and \ref{fig:cos} show the resulting 
momentum and $\cos \theta$ distributions, respectively, 
for charged particles in the laboratory frame.
Very good agreement with MC is found,
except for low-momentum electrons (Fig.~\ref{fig:mom}(a)) and 
pions (Fig.~\ref{fig:mom}(c)).
The dip in the $\cos \theta$ distribution of the muon (Fig.~\ref{fig:cos}(b))
is due to an efficiency drop at the region of overlap 
between the barrel and endcap KLM elements.

\begin{figure}[t]
\centerline{\resizebox{5cm}{5cm}{\includegraphics{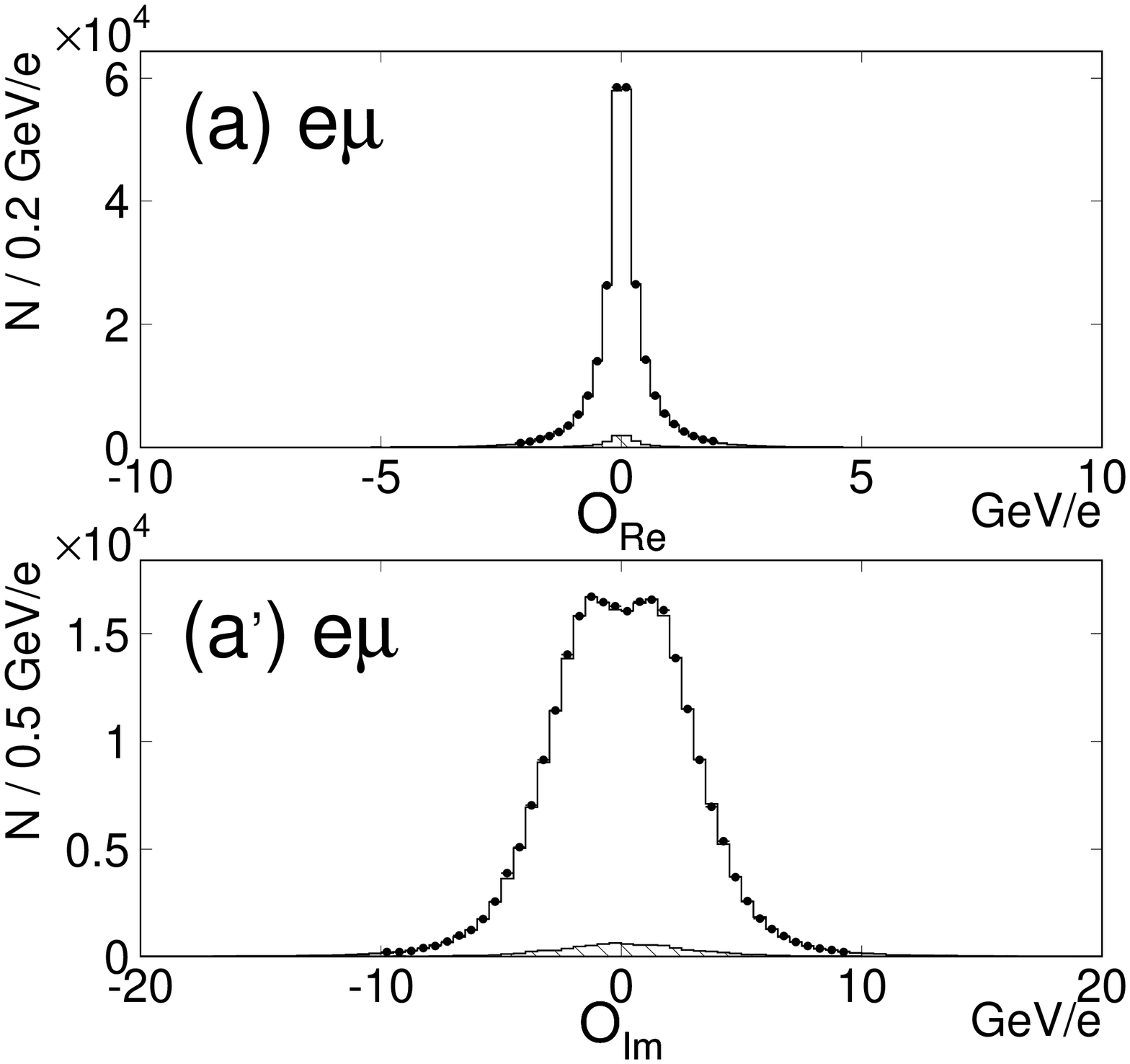}}
            \resizebox{5cm}{5cm}{\includegraphics{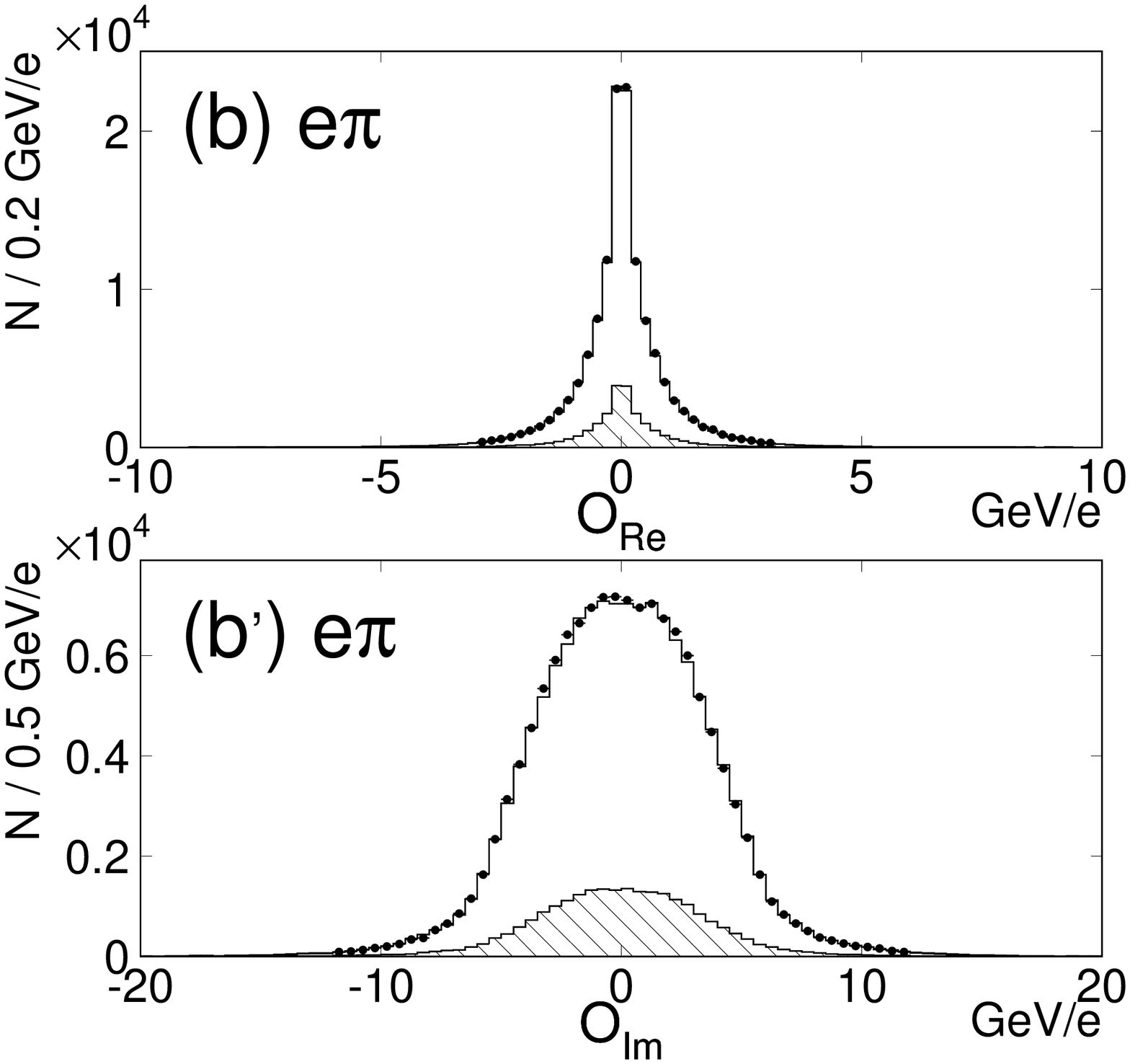}}
            \resizebox{5cm}{5cm}{\includegraphics{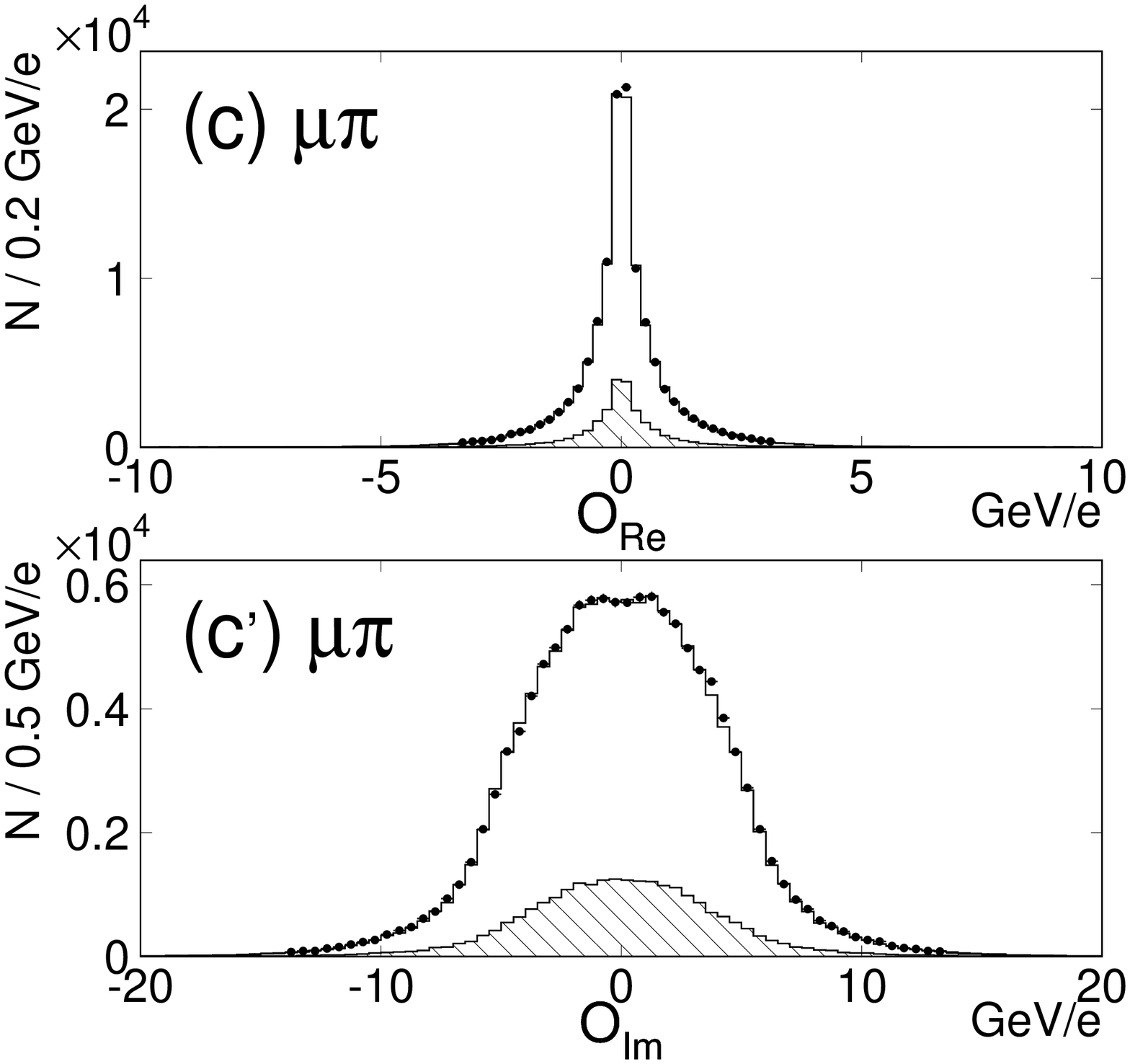}} }
\centerline{\resizebox{5cm}{5cm}{\includegraphics{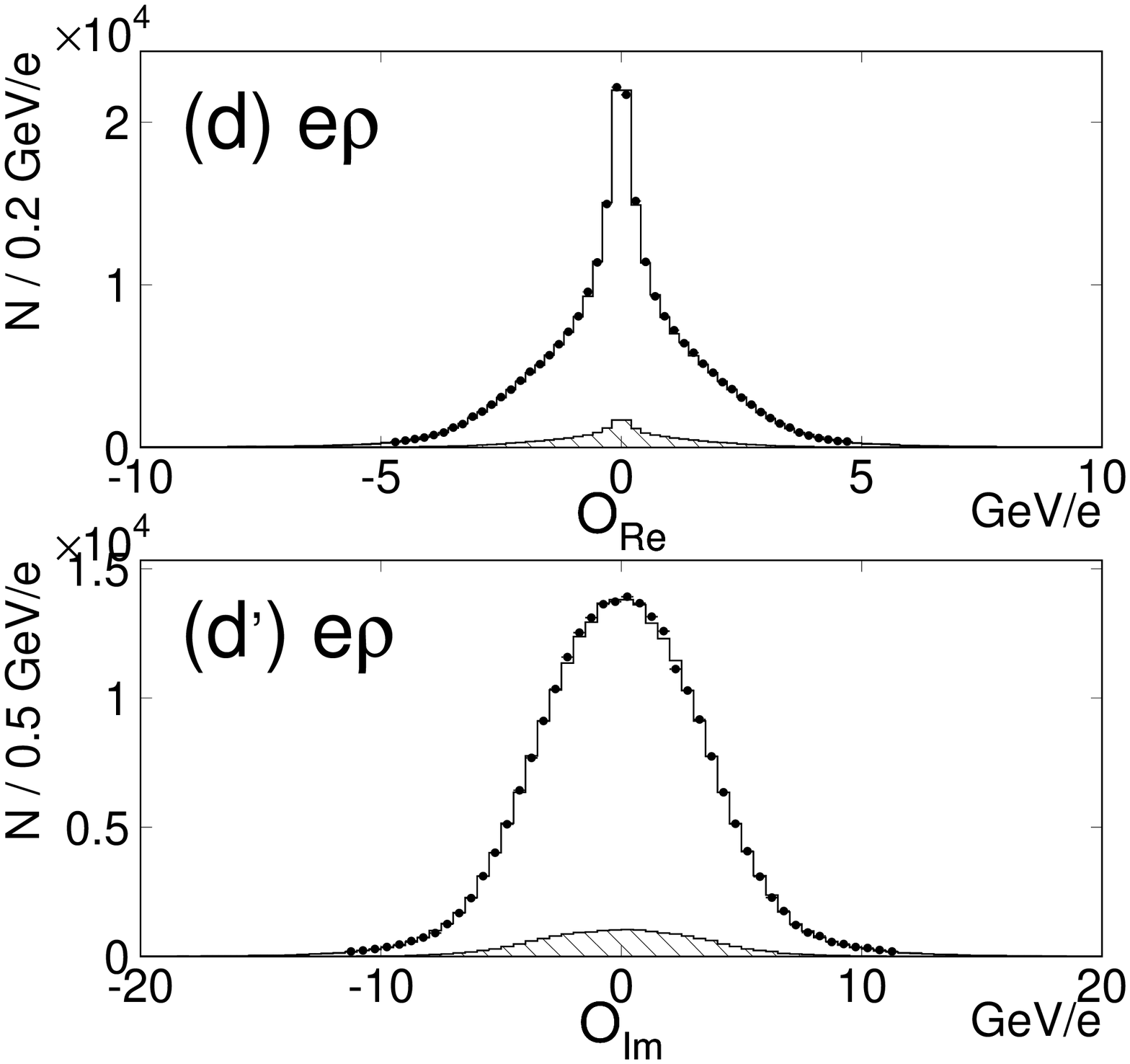}}
            \resizebox{5cm}{5cm}{\includegraphics{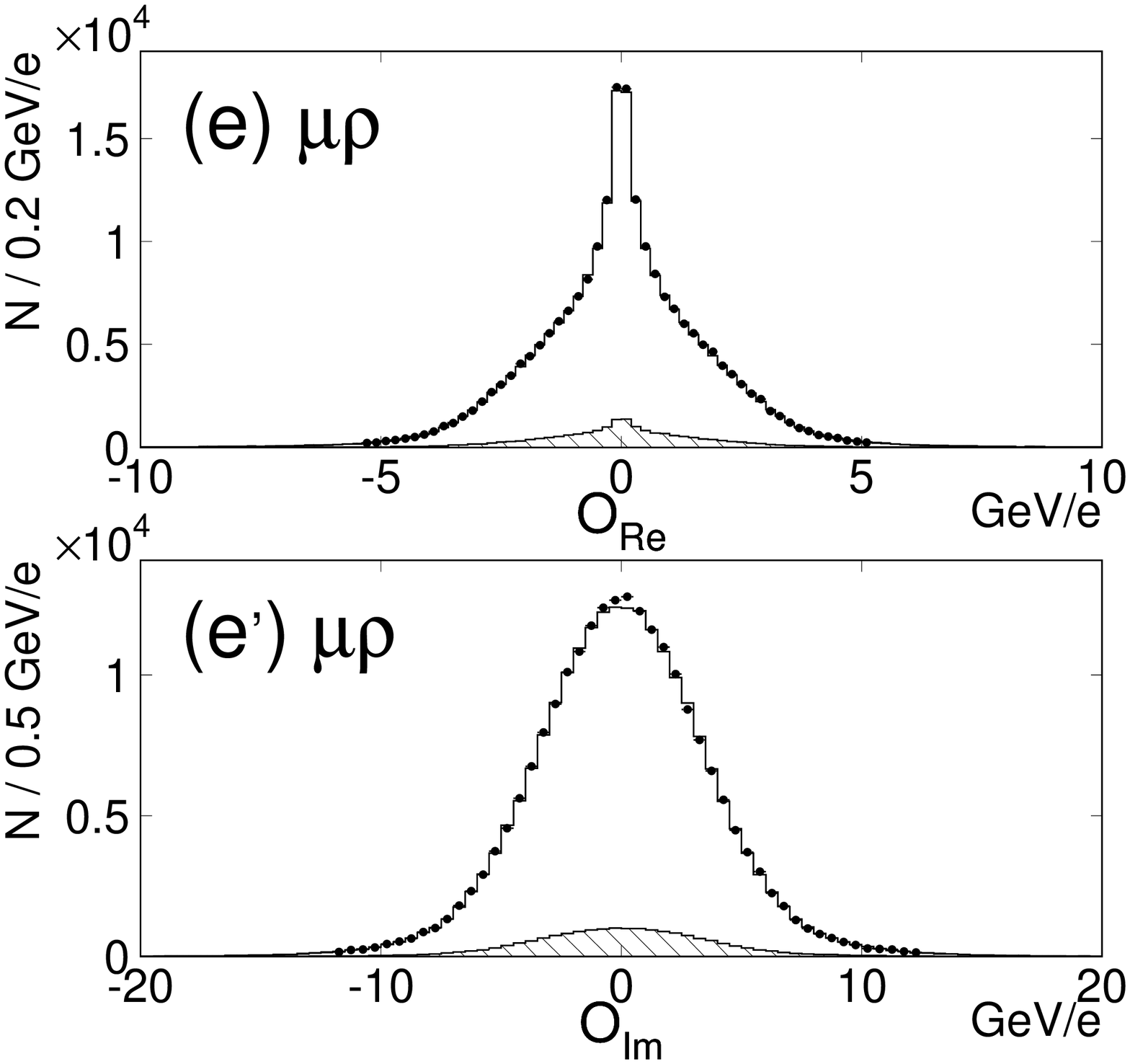}}
            \resizebox{5cm}{5cm}{\includegraphics{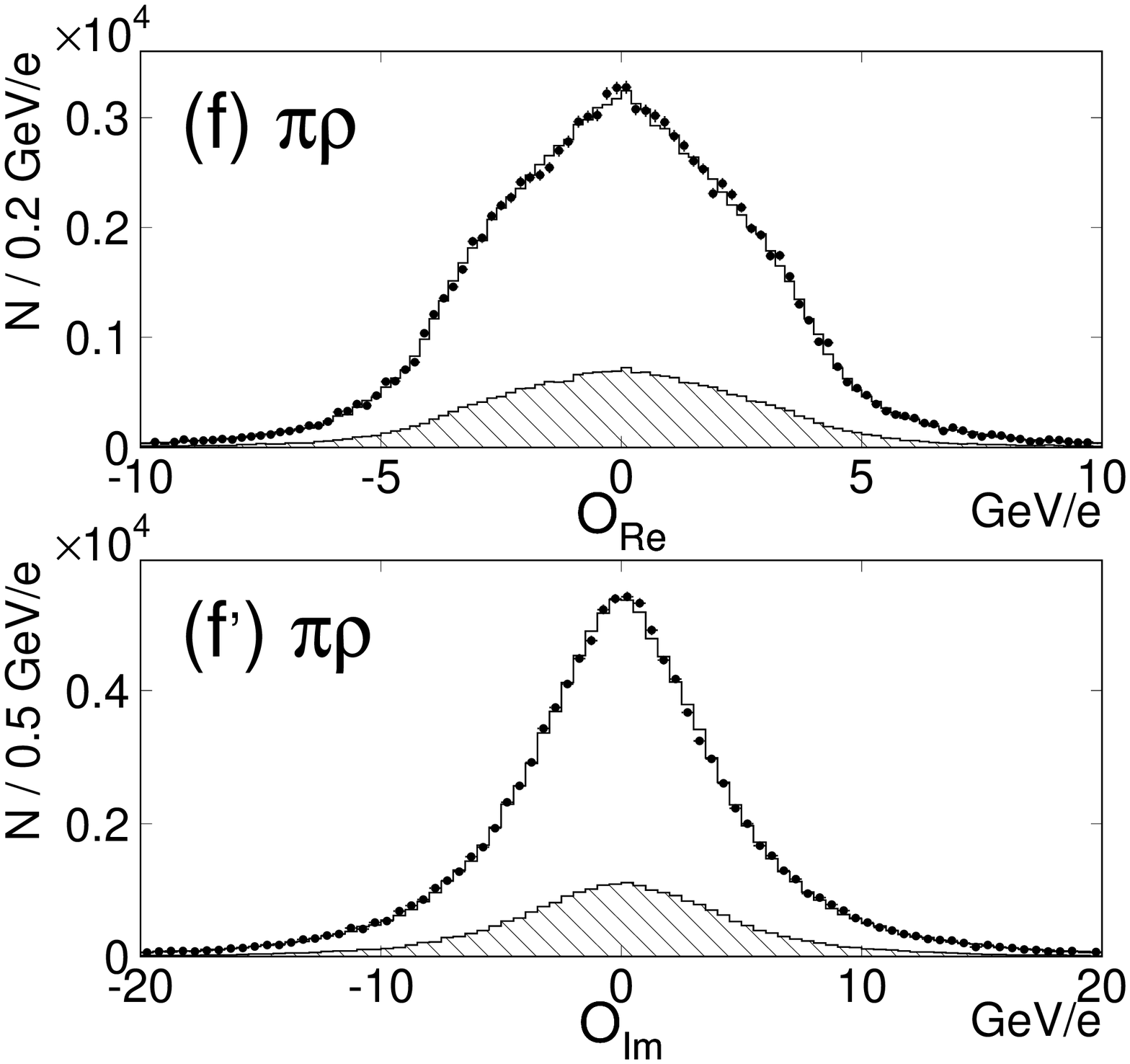}} }
\centerline{\resizebox{5cm}{5cm}{\includegraphics{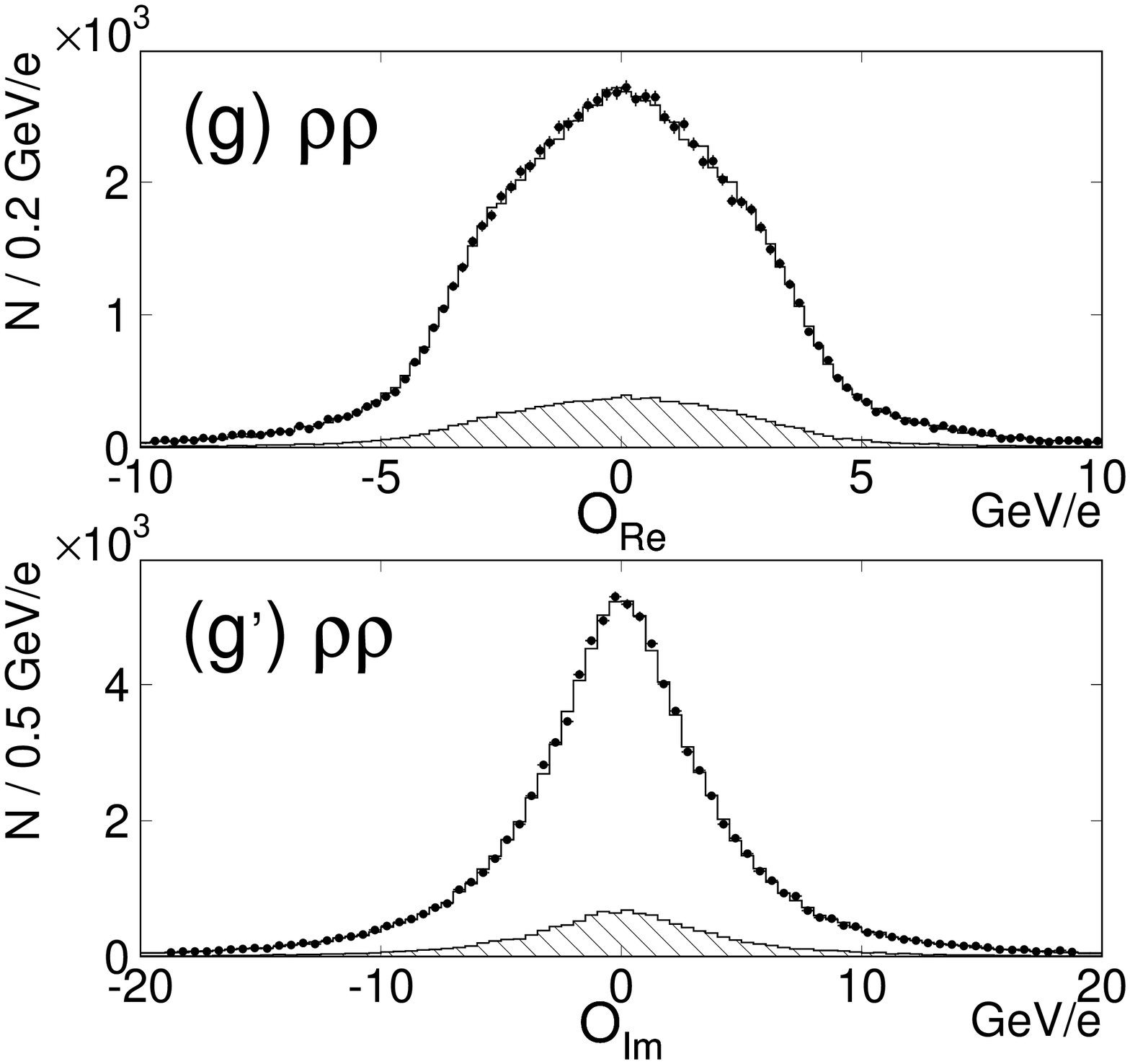}}
            \resizebox{5cm}{5cm}{\includegraphics{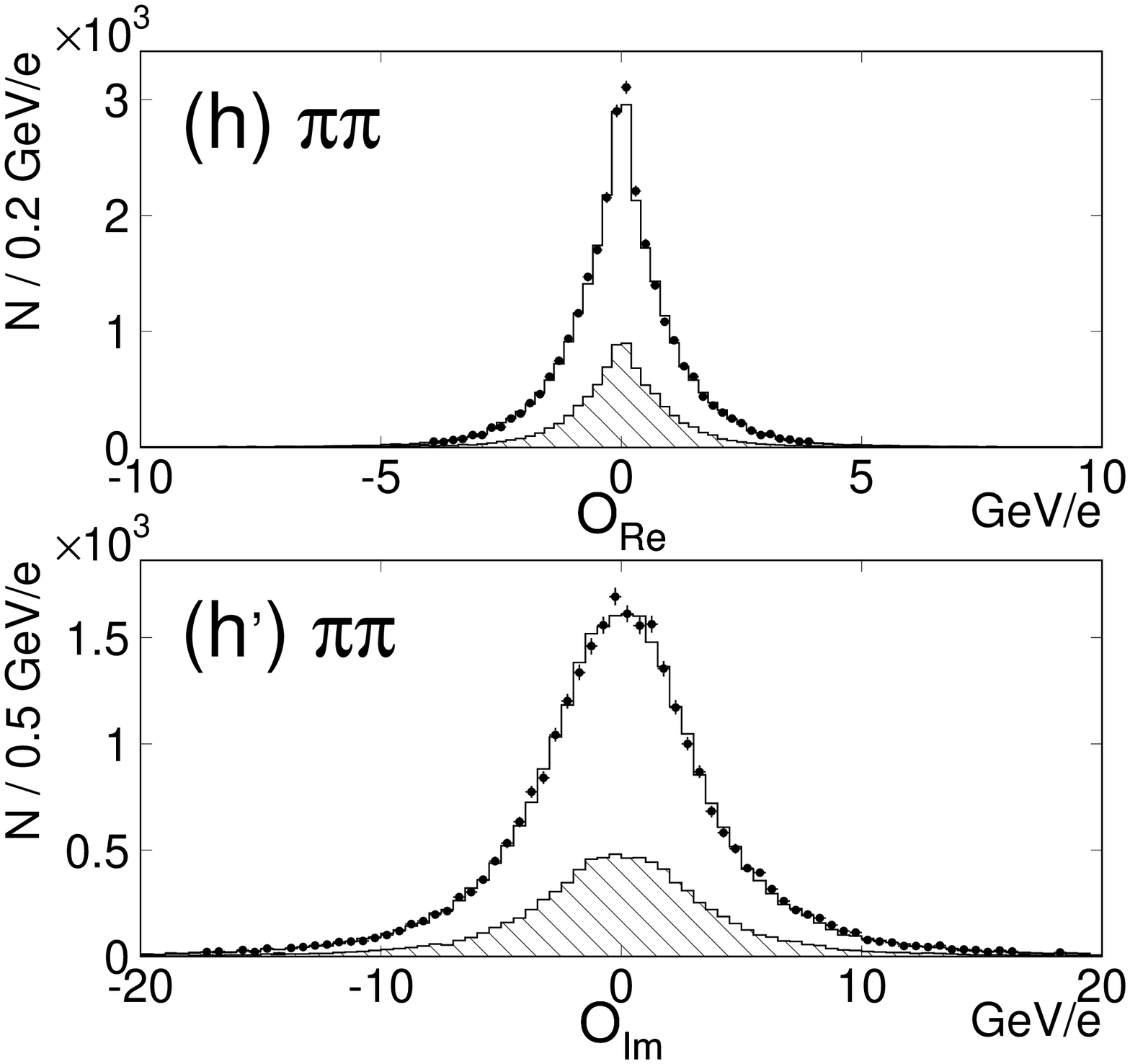}}
            \hspace*{5cm} }
\renewcommand {\baselinestretch}{0.75}
\caption{\small Distributions of the optimal observables 
${\cal O}_{Re}$ and ${\cal O}_{Im}$ for each mode. 
The upper figure of each mode is for ${\cal O}_{Re}$ 
and lower figure is for ${\cal O}_{Im}$.
The closed circles are the experimental data and the histogram is the MC 
expectation with $d_\tau=0$, normalized to the number of entries.
The hatched histogram is the background distribution evaluated by MC.}
\label{fig:Odistribution}
\end{figure}

The resulting ${\cal O}_{Re}$ and ${\cal O}_{Im}$ distributions
are shown in Fig.~\ref{fig:Odistribution} along
with those obtained from MC simulation with $d_\tau = 0$.
Good agreement is found between the experimental data
and the MC samples.
Distributions of the ratio of the data to MC (not shown here)
are flat and near 1.0.

\section{Extraction of $d_\tau$} \label{sec:extraction}

In order to extract the $d_\tau$ value from the observable 
using Eq.~(\ref{eq:relation1}), we have to know the coefficient $a$ and 
the offset $b$.
In the ARGUS analysis~\cite{ref:ARGUS}, which also used the optimal observable
method, the first term
of Eq.~(\ref{eq:obs}) was assumed to be negligible
because it vanishes when the integration takes place over the full phase space: 
$\int {\cal{M}}^2_{Re} d\phi = 0$.
The value of $d_{\tau}$ was obtained as 
the ratio of the observable's mean to the second term,
$Re(d_{\tau})=\langle{\cal O}_{Re}\rangle/\langle{\cal O}^2_{Re}\rangle$.  
However, the detector acceptance $\eta$ affects the
means of the observables according to
\begin{equation}
\langle{\cal {O}}_{Re}\rangle \propto
\int \eta(\phi) {\cal O}_{Re} {\cal M}^2_{\rm prod} d\phi.
\end{equation}
Similar expressions obtain for the imaginary part.
This means that the first term of Eq.~(\ref{eq:obs}) is not necessarily zero
and the coefficient may differ from $\langle{\cal O}^2_{Re}\rangle$
when the detector acceptance is taken into account.
In the ARGUS study, the acceptance effect produced 
the largest systematic uncertainty, of the order of $10^{-16} e$\,cm.

\begin{figure*}[t]
\centerline{\resizebox{7cm}{5.25cm}{\includegraphics{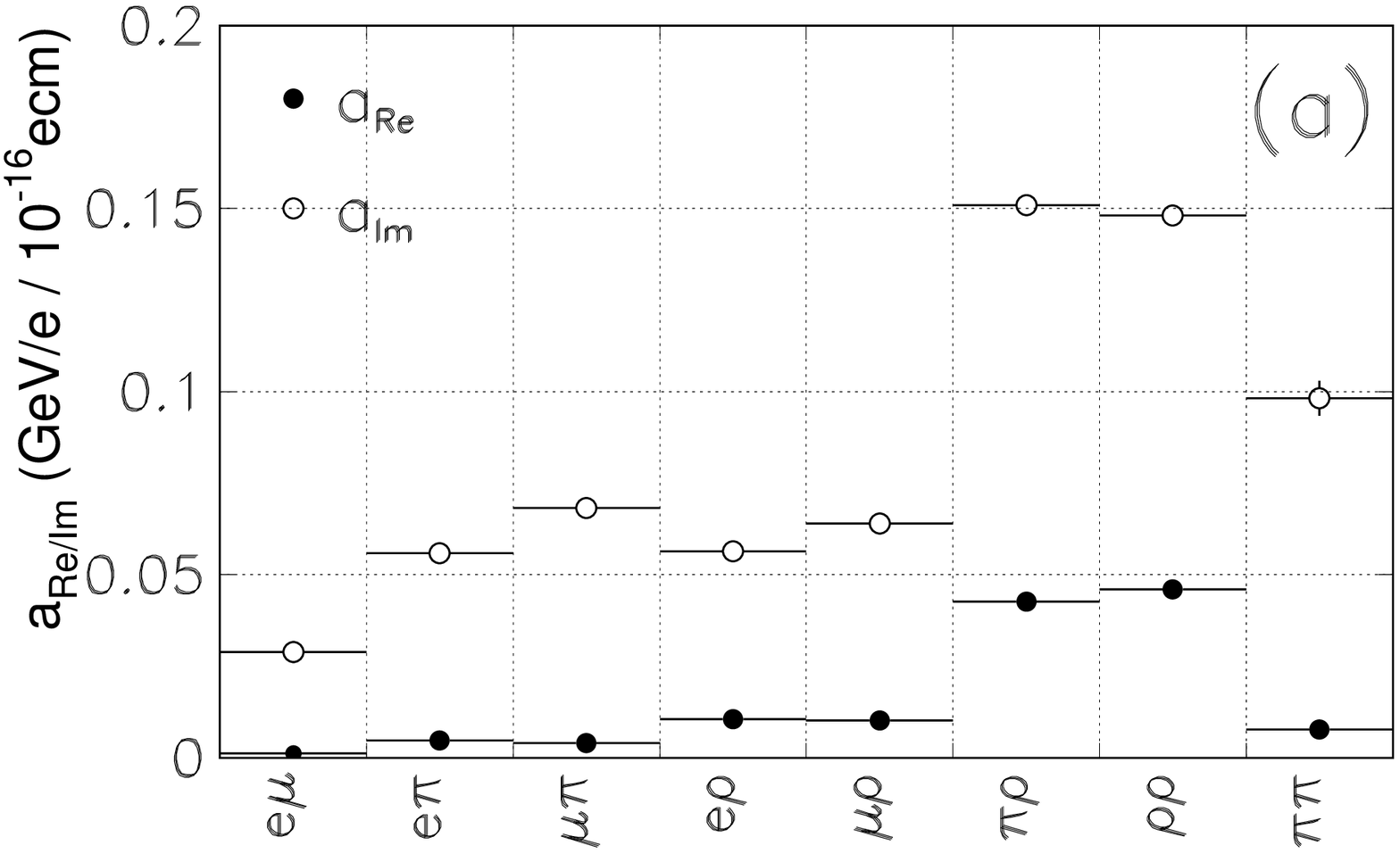}}
            \resizebox{7cm}{5.25cm}{\includegraphics{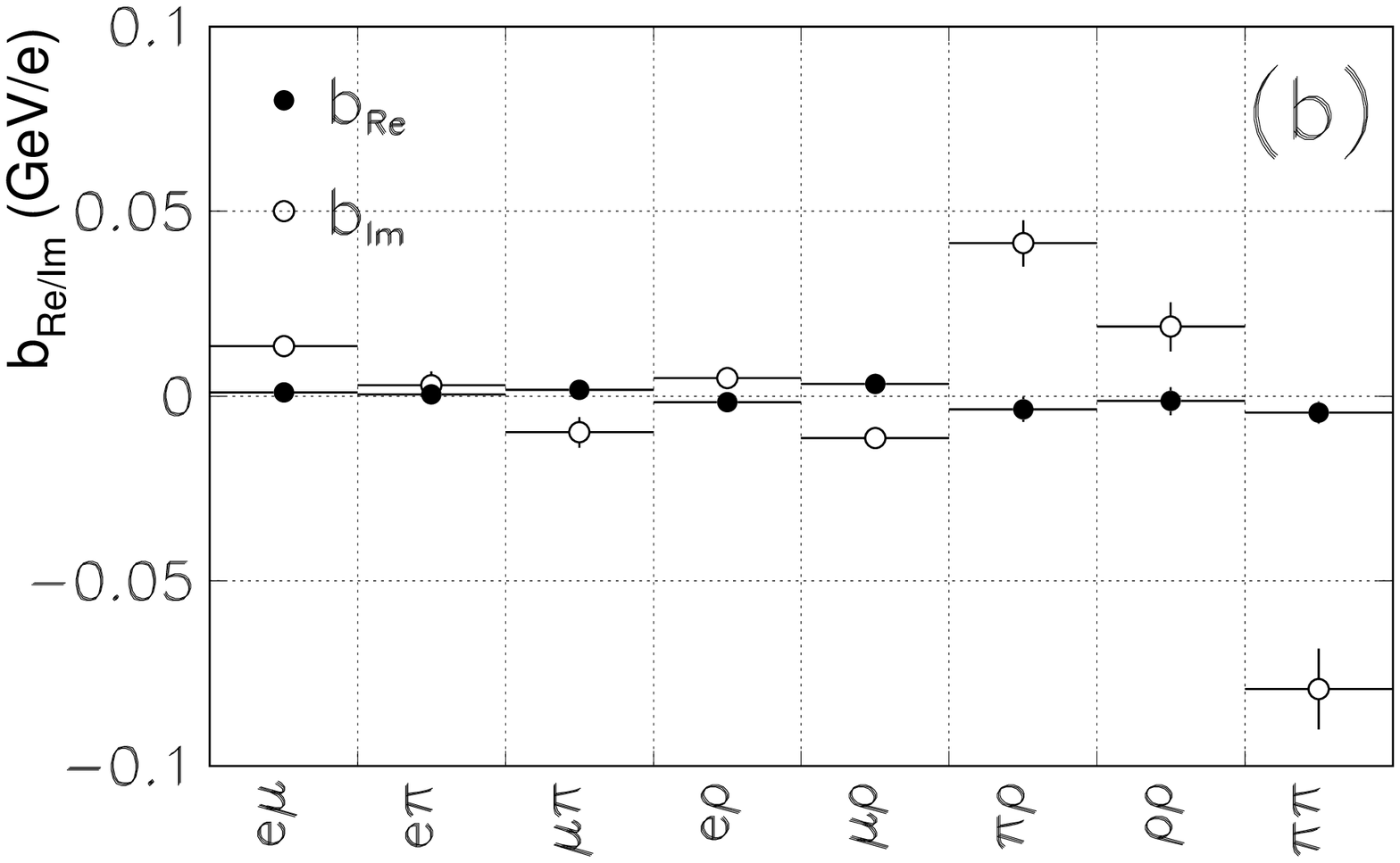}}}
\renewcommand {\baselinestretch}{0.75}
\caption{\small Sensitivity $a$ (a) and 
offset $b$ (b) for each mode from MC. 
The closed circles show the parameters for $Re(d_\tau)$
and the open circles show the parameters for $Im(d_\tau)$.
The errors are due to MC statistics.
}
\label{fig:sensitivity}
\end{figure*}

In order to reduce this systematic effect,
we extract both parameters $a$ and $b$ from the correlation between
$\langle {\cal O}_{Re} \rangle$($\langle {\cal O}_{Im} \rangle$)
and $Re(d_\tau)$($Im(d_\tau)$)
obtained by a full MC including acceptance effects.
\begin{figure*}[t]
\centerline{\resizebox{7cm}{7cm}{\includegraphics{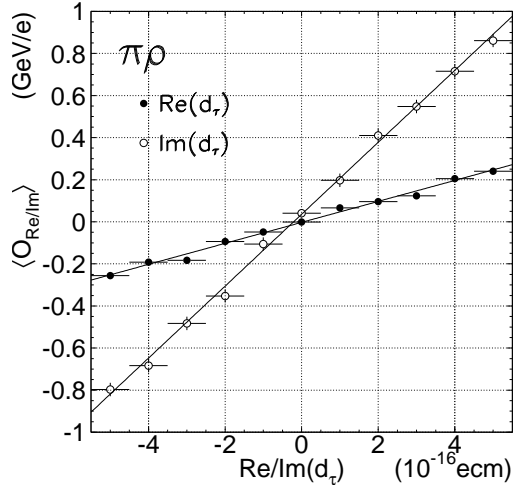}}}
\renewcommand {\baselinestretch}{0.75}
\caption{\small $Re(d_\tau)(Im(d_\tau))$ dependence of the mean of the 
observable $\langle{\cal O}_{Re}\rangle$($\langle{\cal O}_{Im}\rangle$)
for the $\pi\rho$ mode.
The closed circles show the dependence for $Re(d_\tau)$ and 
the open circles show the dependence for $Im(d_\tau)$.
The lines show the fitted linear functions.}
\label{fig:correlation}
\end{figure*}
An example of the correlation between 
$\langle {\cal O}_{Re} \rangle$($\langle {\cal O}_{Im} \rangle$) 
and $Re(d_\tau)$($Im(d_\tau)$) 
is shown in Fig.~\ref{fig:correlation}. Each point
is obtained from MC with detector simulation and event selection. 
By fitting the correlation plot with Eq.~(\ref{eq:relation1}), the parameters 
$a$ and $b$ are obtained.
The feed-across background from other $\tau$ decays
shows some dependence on $d_\tau$, 
because the spin direction is correlated with the momenta of
the final state particles.
Therefore, the effects of the feed-across background on the
coefficient $a$ and offset $b$ are corrected using the parameters
obtained from the background.
The resulting coefficients and offsets are shown
in Fig.~\ref{fig:sensitivity}. 

As can be seen from the values of the coefficient $a$,
the $\pi\rho$ and $\rho\rho$ modes have the highest sensitivities for $d_\tau$,
while the $\pi\pi$ mode has a somewhat lower sensitivity.
For the real part,
this is an effect of averaging over the two solutions for the $\tau$
direction: in the $\tau \to \pi \nu$ case, this causes 
spin correlation information to be lost,
whereas for $\tau \to \rho \nu$
the angular distribution of the $\rho \to \pi \pi^0$ decay
provides information on the $\tau$ spin which survives the
averaging procedure. 
For the imaginary part, the lower sensitivity is due to the 
tighter $\cos \theta$ cut applied to pions
in the $\tau \to \pi \nu$ channel, compared to $\tau \to \rho \nu$;
there is also a small effect from the tighter momentum cut.
The remaining modes, which include
leptons, have low sensitivities because information about
the $\tau$ spin, and also its direction, is lost due to the
additional neutrino(s). 
Non-zero offsets $b_{Im}$ are seen for the imaginary part,
due to the forward/backward asymmetry in the acceptance of the detector.

\section{Systematic uncertainties}

Although in general the MC simulation reproduces the observed 
kinematic distributions well, the small disagreement that is evident
in Figures~\ref{fig:mom} and \ref{fig:cos} dominates
the systematic uncertainty.
The effect on $d_\tau$ is studied by reweighting 
the MC distributions by the ratio of data and MC.
The second significant uncertainty originates from possible charge asymmetry
in the detection efficiency. 
The ratio of yields, $N(\alpha^+\beta^-)/N(\alpha^-\beta^+)$,
for data and MC agrees within 1\%,
where $\alpha$ and $\beta$ are the relevant charged particles
from the $\tau$ decays. 
The resulting systematic uncertainty is evaluated by varying the detection 
efficiency by $\pm1$\%. 
The effect is of the same size as the statistical error for $Im(d_{\tau})$,
while it is negligible for $Re(d_{\tau})$. 
The backgrounds lead to additional systematic uncertainties in $d_\tau$ 
because the parameters $a$ and $b$ are corrected for
the background distributions:
this effect is assessed by varying the assumed background rate.
The effects of photon energy resolution, and of possible
biases in reconstructed momentum (for charged tracks and
photons), are checked by applying scaling factors based on
a comparison of data and MC distributions.
In order to examine a possible asymmetry arising from 
the alignment of the tracking devices,
we measure the differences in polar angles, 
${\it \Delta} \theta = \theta^+_{\rm CM} - \theta^-_{\rm CM}$,
and azimuthal angles, 
${\it \Delta} \phi = \phi^+_{\rm CM} - \phi^-_{\rm CM}$,
for the two tracks in $e^+e^- \to \mu^+\mu^-$ events,
and find a small deviation from back-to-back topological alignment
in each direction: 
${\it \Delta} \theta = 1.48$ mrad and ${\it \Delta} \phi = 0.36$ mrad.
Applying an artificial angular deviation of this magnitude 
to one of the charged tracks, 
we find the change in the observables to be 
negligible compared to the other errors.

To estimate the effect of ignoring radiative processes
in the calculation of the observables, we consider the
case $d_\tau = 0$ and compare our standard calculation to one
which includes initial state radiation, taking the shift
in $d_\tau$ as a measure of the systematic error. Since this
shift occurs in analysis of both data and MC events, 
it is already taken into account in the analysis
(up to effects of detector and/or background mismodelling), 
so the estimate is conservative. For all
decay modes apart from $\pi\pi$, the shift is negligibly
small.

The various sources of systematic error are listed in 
Table~\ref{table:Syserror}.

\begin{table}
\renewcommand {\baselinestretch}{0.75}
 \caption{Systematic errors for $Re(d_\tau)$ and $Im(d_\tau)$
in units of $10^{-16}e\,{\rm cm}$.}
 \label{table:Syserror}
 \begin{center}
 \begin{tabular}{lcccccccc}
  \hline
$Re(d_\tau)$ &$e\mu$&$e\pi$&$\mu\pi$&$e\rho$&$\mu\rho$&$\pi\rho$&$\rho\rho$&$\pi\pi$\\
  \hline
Mismatch of distribution & 0.80 & 0.58 & 0.70 & 0.11 & 0.15 & 0.21 & 0.16 & 0.06 \\
Charge asymmetry         & 0.00 & 0.01 & 0.01 & 0.01 & 0.01 & 0.01 & -    & -    \\
Background variation     & 0.43 & 0.12 & 0.07 & 0.07 & 0.08 & 0.03 & 0.04 & 0.05 \\
Momentum reconstruction  & 0.16 & 0.09 & 0.24 & 0.04 & 0.06 & 0.06 & 0.04 & 0.45 \\
Detector alignment       & 0.02 & 0.02 & 0.01 & 0.00 & 0.01 & 0.01 & 0.02 & 0.03 \\
Radiative effects        & 0.09 & 0.04 & 0.02 & 0.01 & 0.01 & 0.02 & 0.00 & 0.16 \\
  \hline
Total                    & 0.93 & 0.60 & 0.74 & 0.14 & 0.18 & 0.22 & 0.17 & 0.48 \\
  \hline \hline
$Im(d_\tau)$ &$e\mu$&$e\pi$&$\mu\pi$&$e\rho$&$\mu\rho$&$\pi\rho$&$\rho\rho$&$\pi\pi$\\
  \hline
Mismatch of distribution & 0.43 & 0.02 & 0.05 & 0.12 & 0.01 & 0.05 & 0.10 & 0.41 \\
Charge asymmetry         & 0.13 & 0.44 & 0.43 & 0.02 & 0.09 & 0.15 & -    & -    \\
Background variation     & 0.08 & 0.07 & 0.02 & 0.01 & 0.03 & 0.02 & 0.03 & 0.06 \\
Momentum reconstruction  & 0.03 & 0.03 & 0.06 & 0.00 & 0.02 & 0.02 & 0.04 & 0.04 \\
Detector alignment       & 0.01 & 0.03 & 0.02 & 0.01 & 0.01 & 0.02 & 0.01 & 0.05 \\
Radiative effects        & 0.05 & 0.02 & 0.01 & 0.01 & 0.01 & 0.02 & 0.01 & 0.02 \\
  \hline
Total                    & 0.46 & 0.45 & 0.44 & 0.13 & 0.10 & 0.16 & 0.11 & 0.42 \\
  \hline
 \end{tabular}
 \end{center}
\end{table}

\section{Result}

The values of $d_\tau$ extracted using Eq.~(\ref{eq:relation1}) are listed in 
Table~\ref{table:EDMresult} along with the corresponding
statistical and systematic errors,
and plotted in Fig.~\ref{fig:EDMresult}. 
The results are consistent with $d_\tau = 0$ within the errors.

\begin{table}
\renewcommand {\baselinestretch}{0.75}
 \caption{Results for the electric dipole moment. 
The first error is statistical and the second is systematic.}
 \label{table:EDMresult}
 \begin{center}
 \begin{tabular}{ccc}
  \hline
  Mode       & $Re(d_\tau)~~(10^{-16}e\,{\rm cm})$&$Im(d_\tau)~~(10^{-16}e\,{\rm cm})$\\
  \hline
      $e\mu$ & $~~2.25 \pm 1.26 \pm 0.93$ & $ -0.41 \pm 0.22 \pm 0.46$ \\
      $e\pi$ & $~~0.43 \pm 0.64 \pm 0.60$ & $ -0.22 \pm 0.19 \pm 0.45$ \\
    $\mu\pi$ & $ -0.41 \pm 0.87 \pm 0.74$ & $~~0.15 \pm 0.19 \pm 0.44$ \\
     $e\rho$ & $~~0.00 \pm 0.36 \pm 0.14$ & $ -0.01 \pm 0.14 \pm 0.13$ \\
   $\mu\rho$ & $~~0.04 \pm 0.42 \pm 0.18$ & $ -0.02 \pm 0.14 \pm 0.10$ \\
   $\pi\rho$ & $~~0.34 \pm 0.25 \pm 0.22$ & $ -0.22 \pm 0.13 \pm 0.16$ \\
  $\rho\rho$ & $ -0.08 \pm 0.25 \pm 0.17$ & $ -0.12 \pm 0.14 \pm 0.11$ \\
    $\pi\pi$ & $~~0.42 \pm 1.17 \pm 0.48$ & $~~0.24 \pm 0.34 \pm 0.42$ \\
  \hline
  Mean value & $0.115 \pm 0.170$          & $-0.083 \pm 0.086$ \\
  \hline
 \end{tabular}
 \end{center}
\end{table}
\begin{figure}
\centerline{\resizebox{7cm}{5.25cm}{\includegraphics{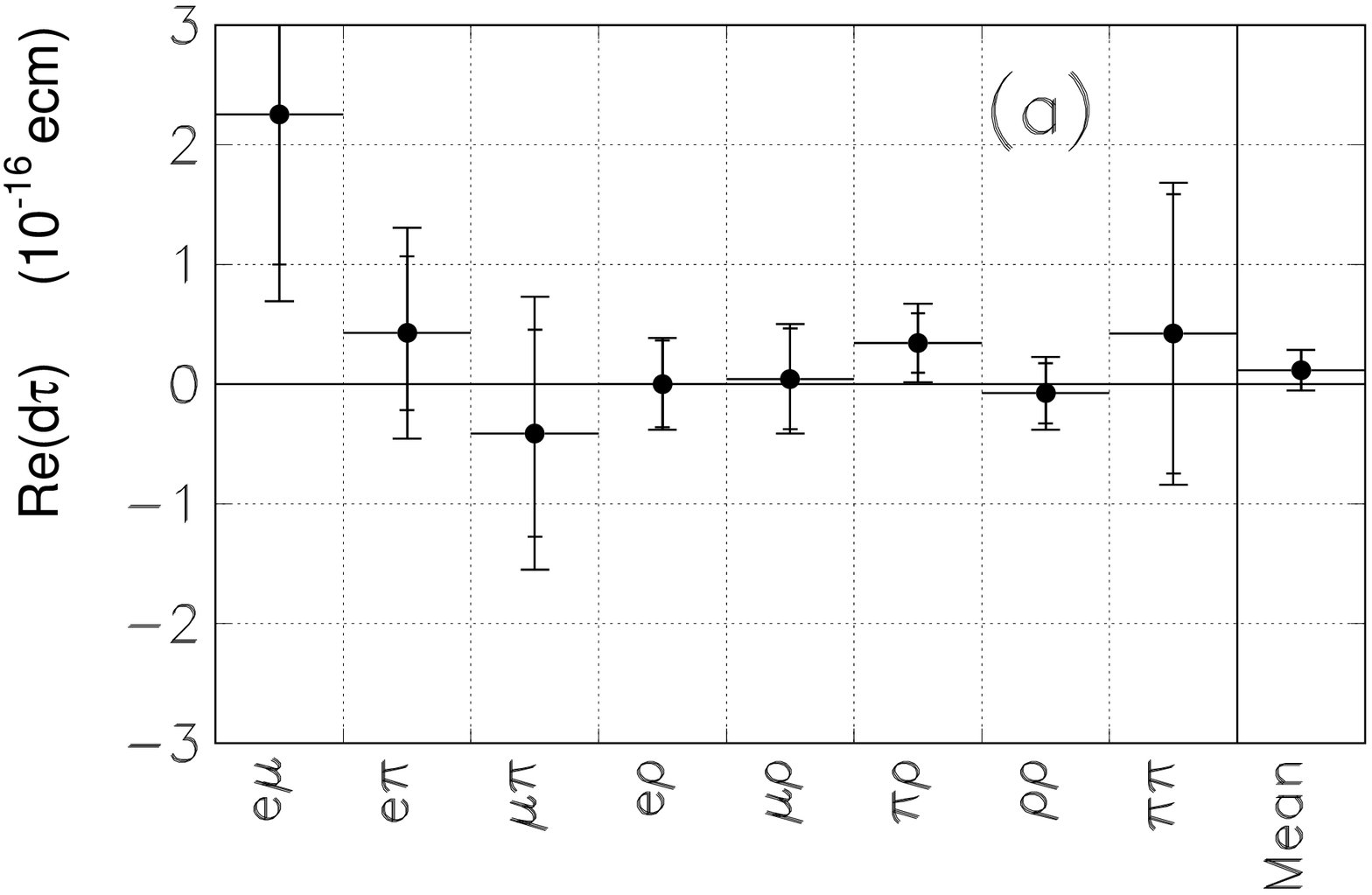}}
            \resizebox{7cm}{5.25cm}{\includegraphics{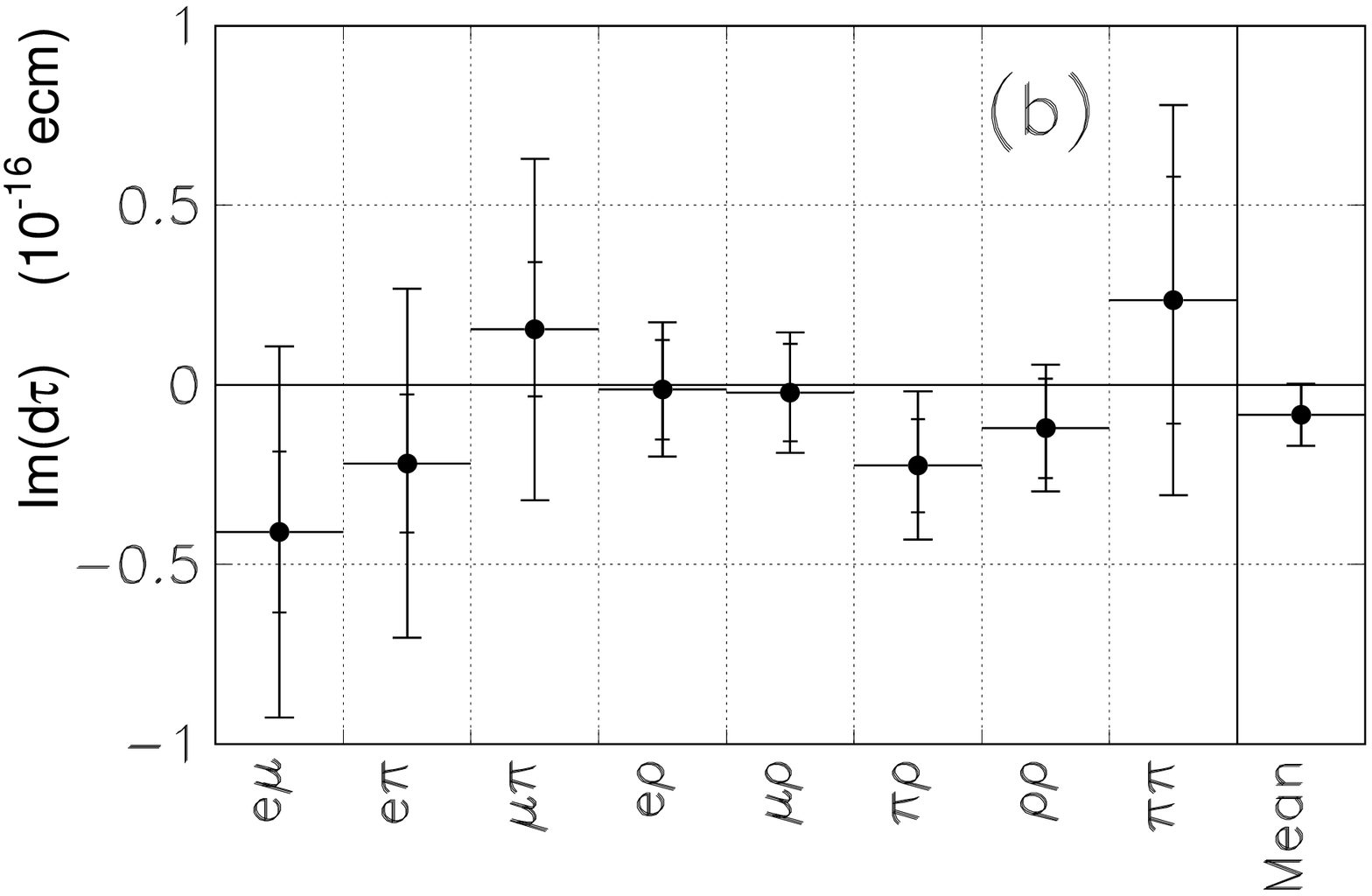}}}
\renewcommand {\baselinestretch}{0.75}
\caption{\small $Re(d_\tau)$ and $Im(d_\tau)$ for each mode.
Both statistical and systematic errors are included.
The small ticks on the error bars show the statistical errors.}
\label{fig:EDMresult}
\end{figure}

Finally, we obtain mean values for $Re(d_\tau)$ and $Im(d_\tau)$
over the eight different $\tau^+\tau^-$ modes weighted by quadratically
summed statistical and systematic errors,
\begin{eqnarray}
 Re(d_\tau) &=& ( 1.15 \pm 1.70 ) \times 10^{-17} e\,{\rm cm}, \\
 Im(d_\tau) &=& ( -0.83 \pm 0.86 ) \times 10^{-17} e\,{\rm cm},
\end{eqnarray}
with corresponding 95\% confidence limits
\begin{eqnarray}
-2.2 < Re(d_\tau) < 4.5 ~~~(10^{-17} e\,{\rm cm}), \\
-2.5 < Im(d_\tau) < 0.8 ~~~(10^{-17} e\,{\rm cm}).
\end{eqnarray}

This investigation has improved the sensitivity to the $\tau$ lepton's
electric dipole moment by an order of magnitude over
previous measurements. 

\smallskip
\bigskip
\noindent
{\bf Acknowledgements}
\smallskip

We would like to thank Professors K.~Hagiwara, O.~Nachtmann, and Z.~W\c{a}s 
for their constructive advice and many helpful discussions.
We also wish to thank the KEKB accelerator group for the excellent
operation of the KEKB accelerator.
We gratefully acknowledge the support from the Ministry of Education,
Culture, Sports, Science, and Technology of Japan,
Grant-in-Aid for JSPS Fellows 01655 2001,
and the Japan Society for the Promotion of Science;
the Australian Research Council
and the Australian Department of Industry, Science and Resources;
the National Science Foundation of China under contract No.~10175071;
the Department of Science and Technology of India;
the BK21 program of the Ministry of Education of Korea
and the CHEP SRC program of the Korea Science and Engineering Foundation;
the Polish State Committee for Scientific Research
under contract No.~2P03B 17017;
the Ministry of Science and Technology of the Russian Federation;
the Ministry of Education, Science and Sport of the Republic of Slovenia;
the National Science Council and the Ministry of Education of Taiwan;
and the U.S.\ Department of Energy.

\end{document}